\documentstyle[fleqn,psfig,epsf]{elsart}

\newcommand{\be}{\begin{equation}}
\newcommand{\ee}{\end{equation}}
\newcommand{\bea}{\begin{eqnarray}}
\newcommand{\eea}{\end{eqnarray}}
\newcommand \ga{\raisebox{-.5ex}{$\stackrel{>}{\sim}$}}
\newcommand \la{\raisebox{-.5ex}{$\stackrel{<}{\sim}$}}

\newcommand{\av}[1]{\langle{#1}\rangle}

\begin{document}
\begin{frontmatter}

\title{EVENT-BY-EVENT PHYSICS IN\\ 
RELATIVISTIC HEAVY-ION COLLISIONS}

\author{Henning Heiselberg}

\address{NORDITA, Blegdamsvej 17, DK-2100 Copenhagen \O, Denmark}

\maketitle

\begin{abstract}
    Motivated by forthcoming experiments at RHIC and LHC, and results
from SPS, a review is given of the present state of 
event-by-event fluctuations in ultrarelativistic heavy-ion collisions.
Fluctuations in particle multiplicities, ratios, transverse momenta,
rapidity, etc. are calculated in participant nucleon as well as
thermal models. The physical observables, including multiplicity,
kaon to pion ratios, and transverse momenta agree well with recent
NA49 data at the SPS, and indicate that such studies do not yet reveal
the presence of new physics.  Predictions for RHIC and LHC energies
are given. The centrality dependence with and without a phase
transition to a quark-gluon plasma is discussed - in particular, how
the physical quantities are expected to display a qualitative
different behavior in case of a phase transition, and 
can be signaled by anomalous fluctuations
and correlations in a number of observables.

\end{abstract}

\end{frontmatter}

\newpage
\tableofcontents
\newpage
\section{Introduction}

The importance of event-by-event physics is evident from the following
simple analogy: Stick a sheet of paper out of your window on a rainy
day. Keeping it there for a long time - corresponding to averaging -
the paper will become uniformly wet and one would conclude that rain
is a uniform mist. If, however, one keeps the sheet of paper in the
rain for a few seconds only, one observes the striking droplet
structure of rain\footnote{Originally, this analogy was given by
Prof. A.D. Jackson}. 
Incidentally, one has also demonstrated the {\it
liquid-gas phase transition!} Analyzing many events gives good
statistics and may reveal rare events as snow or hail and thus other
phase transitions. The statistics of droplet sizes will also tell
something about the fragmentation, surface tension, etc. By varying
initial conditions as timing and orienting the paper, one can further
determine the speed and direction of the rain drops.

    Central ultrarelativistic collisions at RHIC and LHC are expected
to produce about $\sim 10^4$ particles, and thus present one with
the remarkable opportunity to analyze, on an event-by-event
basis, fluctuations in physical observables such particle
multiplicities, transverse momenta, correlations and ratios.  Analysis
of single events with large statistics can reveal very different
physics than studying averages over a large statistical sample of
events.  
The use of Hanbury Brown--Twiss correlations to extract the
system geometry is a familiar application of event-by-event
physics in nuclear collisions~\cite{HBT} and elsewhere, e.g, in
sonoluminoscence~\cite{Trentalange}. The power of this tool has been
strikingly illustrated in study of interference between Bose-Einstein
condensates in trapped atomic systems \cite{bec}.  
Fluctuations in the
microwave background radiation as recently measured by COBE
\cite{COBE} restrict cosmological parameters for the single Big Bang
event of our Universe.  Large neutrons stars velocities have been
measured recently \cite{kick} which indicate that the supernova
collapse is very asymmetrical and leads to large event-by-event
fluctuations in ``kick'' velocities during formation of neutron stars.

The tools applied to study these phenomena do, however, vary in order to
optimize the analysis and due to limited statistics. The COBE and the
interference in Bose-Einstein condensates require study of
fluctuations within a single event. The HBT studies in heavy-ion
collisions and sonoluminosence requires further averaging over many
events in order to obtain sufficient statistics; one has not yet
studied fluctuations in source radii event-by-event.  Anisotropic flow
requires an event-plane reconstruction in each event \cite{BMS}
but again averaging over many events is necessary to 
obtain a statistically relevant measurement of the flow.
The event-by-event fluctuations in heavy-ion collisions
(and neutron star kick velocities) go a step further by studying
variations from event to event.

    Studying event-by-event fluctuations in ultrarelativistic heavy
ion collisions to extract new physics was proposed in a series of
papers \cite{BFS,PRL91,sigfluct}, in which the analysis of transverse
energy fluctuations in central collisions~\cite{Aa} was used to
extract evidence within the binary collision picture for color, or
cross-section, fluctuations.  More recent theoretical papers have
focussed on different aspects of these fluctuations, such as searching
for evidence for thermalization~\cite{NA49,GazMrow,Mrow}, correlations
between transverse momentum and multiplicity \cite{Trainor}, critical
fluctuations at the QCD phase
transition~\cite{Stodolsky,Shuryak,SRS,Berdnikov,GP} and other
correlations between collective quantities \cite{VKR}.

 Intermittency \cite{Bialas} studies of factorial moments of
multiplicities are related to event-by-event fluctuations.  One of the
motivations for intermittency studies was the idea of self-similarity
on small scales, an idea borrowed from chaos theories.  The factorial
moments of particle multiplicities did find approximate power law
behavior when the intervals of rapidity and angles were made
increasingly smaller, at least until a certain small scale. The power
law scaling in nucleus-nucleus collisions was, however, weaker than
in proton-proton collisions. This indicated that the stronger
correlations in proton-proton collisions were mainly due to
resonances, minijets and other short range correlations, but that they
were averaged out in nuclear collisions by summing over the many
individual participating nucleons. The scaling was not a collective
phenomenon and indications of new physics were not found \cite{Int}.
In more recent event-by-event fluctuation studies the self-similar
scaling idea is abandoned.  They concentrate on the mean and the variance
of the particle multiplicities per event and correlations between different
particle species, transverse momentum, azimuthal angle, etc. One directly
compares to expectations from proton-proton collisions scaled up by
the number of participants.
One follows these fluctuations and correlations for heavy-ion 
collisions as function of centrality and system size searching for
anomalous behavior as compared to proton-proton collisions. 

Recently NA49 has
presented a prototypical event-by-event analysis of fluctuations in
central Pb+Pb collisions at 158 GeV per nucleon at the SPS, which
produce more than a thousand particles per event~\cite{NA49}.
The analysis has been carried out on $\sim$100.000 such events
measuring fluctuations in multiplicities, particle ratios, transverse
momentum, etc.

Results from the RHIC collider are eagerly awaited
\cite{PHOBOS}.  The hope is to observe the phase transition to
quark-gluon plasma, the chirally restored hadronic matter and/or
deconfinement. This may be by distinct signals of enhanced rapidity
and multiplicity fluctuations \cite{BH,SRS} in conjunction with
J/$\Psi$ suppression, strangeness enhancement, $\eta'$ enhancement,
constant (critical) temperatures vs. transverse enery or rapidity density
\cite{HoveT}, transverse flow or other collective
quantities as function of centrality, transverse energy or
multiplicity as will be discussed in detail below.

The purpose of this review is to understand these and other possible
fluctuations. We find that
the physical observables, including multiplicity,
kaon to pion ratios, and transverse momenta agree well with recent
NA49 data at the SPS, and indicate that such studies do not yet reveal
the presence of new physics.  Predictions for RHIC and LHC energies
are given.  The centrality dependence with and without a phase
transition to a quark-gluon plasma is discussed - in particular, how
the physical quantities are expected to display a qualitative
different behavior in case of a phase transition, and how a first
order phase transition could be signaled by very large fluctuations.

\section{Phase Transitions and Fluctuations}

Lattice QCD calculations find a phase transition in strongly
interacting matter 
which is accompanied by a strong increase of the number of effective 
degrees of freedom
\cite{Lattice,BOYD}. The Early Universe underwent 
this transition at a time
$t= 0.3-0.4(T_c/MeV)^{-2}$~seconds. For a
hadronic gas melting temperature of $T_c=150$~MeV this occurred
around $15$~microseconds 
after the Big Bang. By colliding heavy nuclei we
expect to reproduce this transition at sufficiently high collisions
energies.

\subsection{Order of the QCD phase transition}
 
The nature and order of the transition is not known very well.
Lattice calculations can be performed for zero quark and baryon
chemical potential only, $\mu_B=0$, where they suggest that QCD has a
weak first order transition provided that the strange quark is sufficiently
light \cite{Lattice,BOYD}, that is for 3 or more massless quark
flavors. 
The transition is due to chiral symmetry restoration and occur
at a critical temperature $T_C\simeq 150$~MeV.
In pure SU(3) gauge
theory (that is no quarks, $N_f=0$) the transition is a deconfinement
transition which is of
first order and occurs at a higher temperature $T_c\simeq 260$~MeV.

However, when the strange or the up and down quark masses become
massive, the QCD transition changes to a smooth cross over. The phase
diagram is then like the liquid-gas phase diagram with a critical
point above which the transition goes continuously through the vapor
phase.  For reasonable values for the strange quark mass, $m_s\sim
150$~MeV and small up and down quark masses, lattice
calculations find either a weak first order transition 
\cite{Lattice,BOYD} or a smooth soft cross-over \cite{JLQCD}. 
In case of a weak first order transition, the
latent heat and density discontinuities
and the signals, that depend on these quantities, will be small.

For exactly two massless flavors, $m_{u,d}=0$ and $m_s=\infty$, the
transition is second order at small baryon chemical potential. Random
matrix theory finds a 2nd order phase transition at high temperatures
which, however, change into a 1st order transition above a certain
baryon density - the tricritical point.
For small up and down quark masses the transition changes to a continuous
cross-over at zero baryon chemical potential but remains a first order
at large baryon chemical potential. A critical point must therefore exist
at small but finite baryon chemical potential which may be searched for
in relativistic heavy-ion collisions \cite{SRS,Berdnikov,GP}.

\begin{figure}[htb] 
\begin{center} 
\vspace*{2.5cm} 
\centerline{\hspace{-2cm}\epsfxsize=8cm\epsfbox{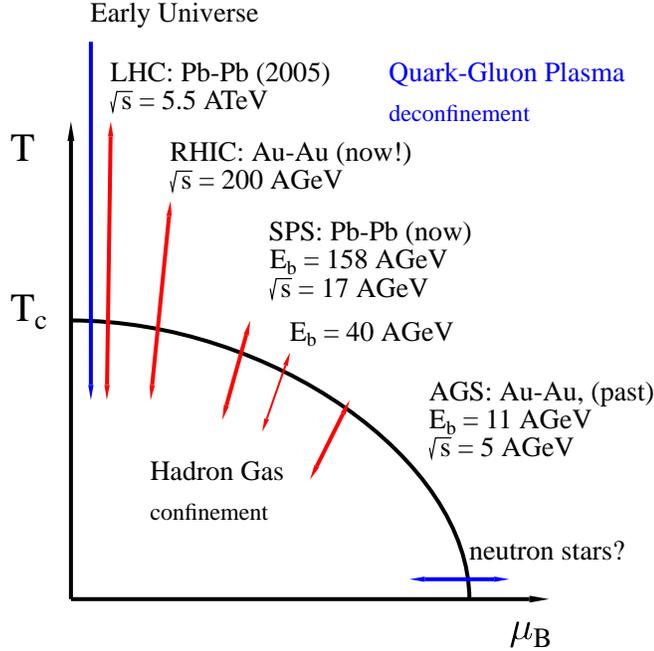}}
\end{center} 
\vspace*{-2cm} 
\caption{An illustration of the QCD phase diagram, temperature
vs. baryon chemical potential. The regions of the phase diagram
probed by various high energy nuclear collisions are sketched
by arrows. From \protect\cite{Eskola}.}
\label{Eskola}
\end{figure} 

\subsection{Density, rapidity, temperature and other fluctuations}

\begin{figure}[t]
\vskip -.5cm
\psfig{figure=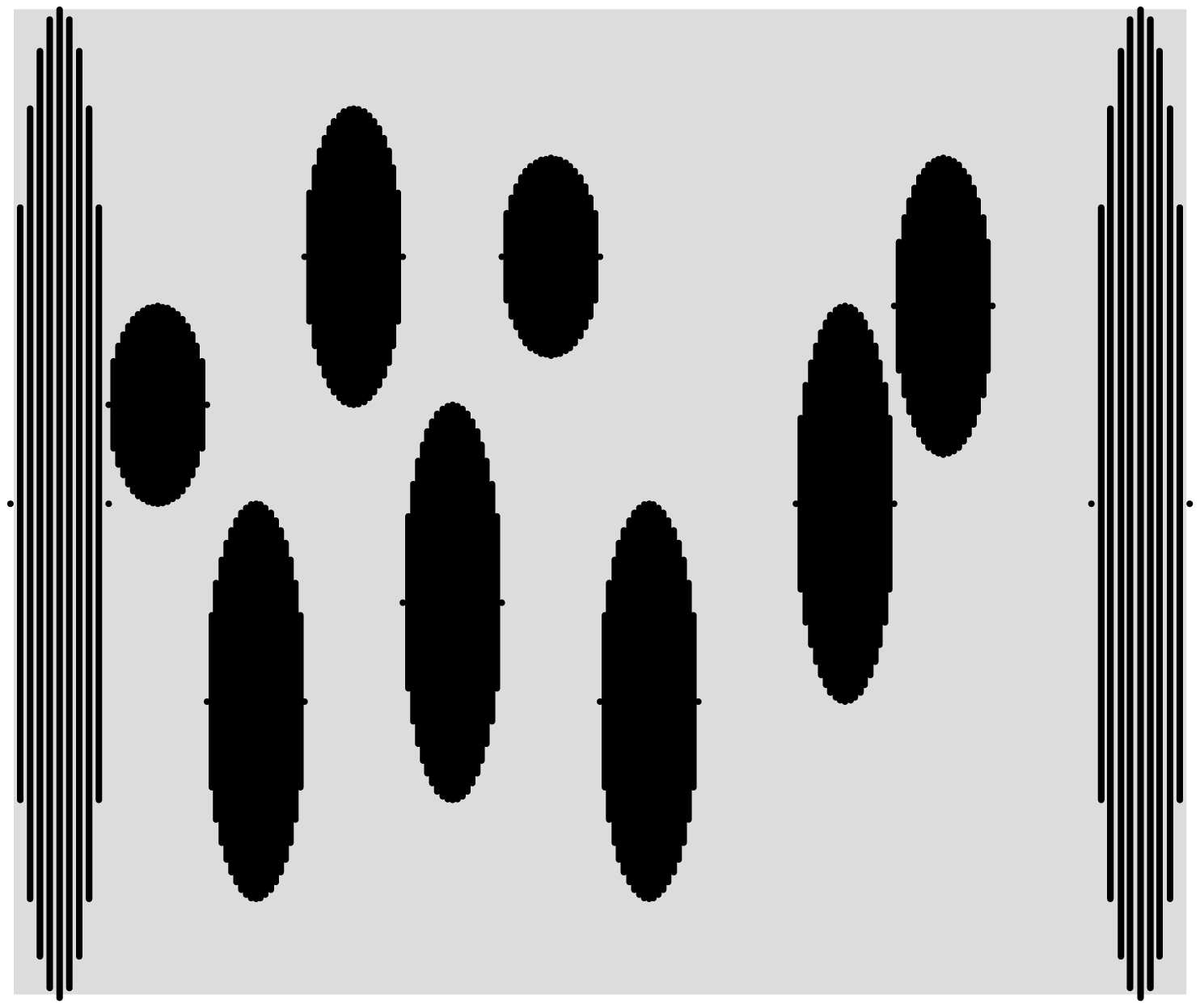,height=3.5in,angle=0}
\vspace{-1.8cm}
\psfig{figure=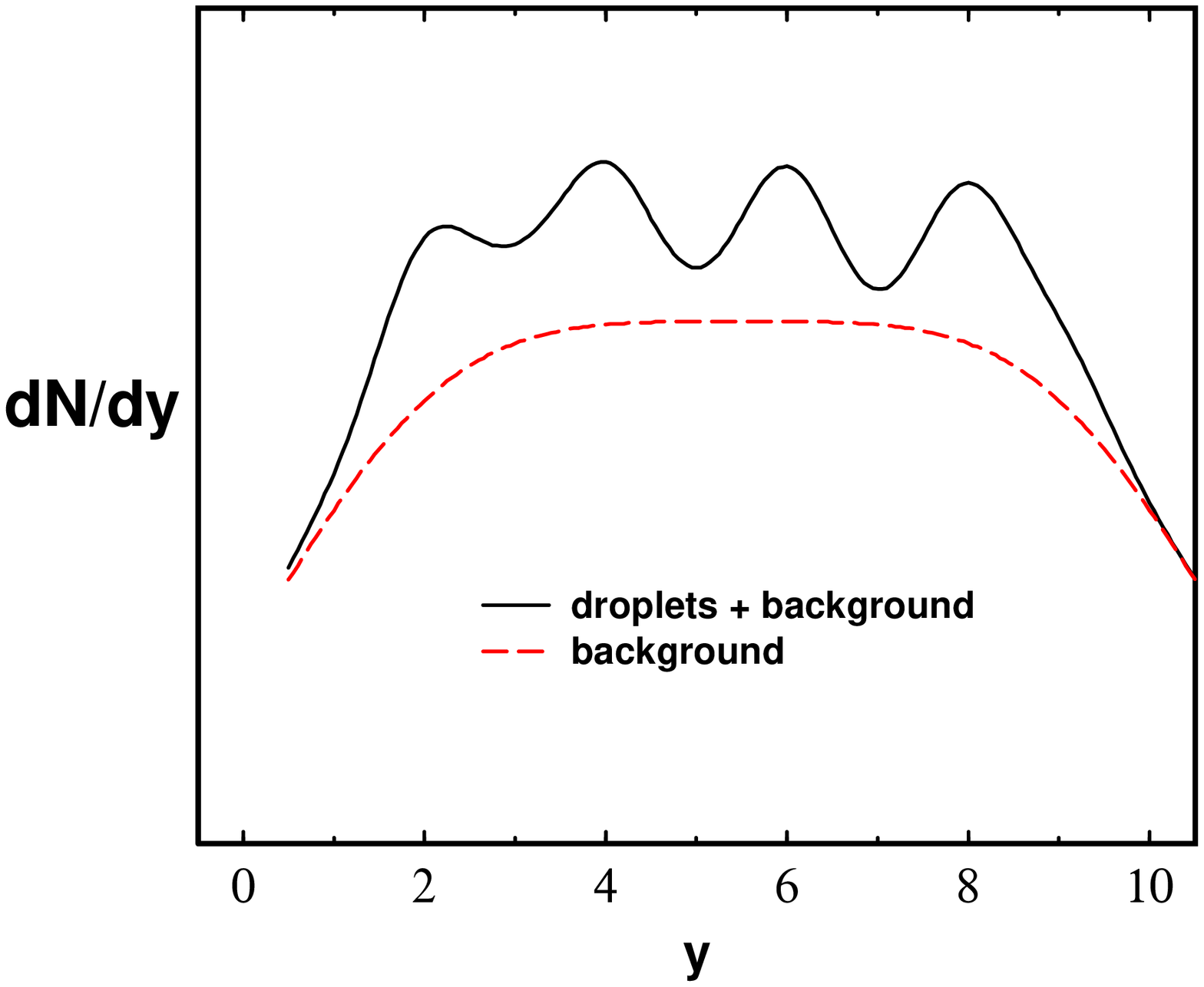,height=3.5in,angle=0}
\caption{Sketch of droplet formation (top) in a continuous background
of hadron vs. $\eta$. The corresponding
rapidity distributions (bottom) are shown for the continuous 
hadronic background with and without droplets
(target and projectile fragmentation regions are excluded).  \label{drop}}
\end{figure}

Fluctuations are very sensitive to the nature of the transition.
In case of a second order phase transition the specific heat
diverges, and this has been argued to reduce the fluctuations drastically
if the matter freezes out at $T_c$ \cite{Stodolsky,Shuryak,SRS,Berdnikov}.
For example, the temperature fluctuations have a probability 
distribution \cite{Landau}
\bea
  w \sim \exp(-C_V (\frac{\Delta T}{T})^2) \,;
\eea
a diverging specific heat near a 2nd order phase transition would
then remove fluctuations if matter is in global thermal equilibrium.
The implications of such critical phenomena 
near second order phase transitions
and critical points are discussed in detail in \cite{SRS,Berdnikov}.
It is found that the expansion of the systems slows the growth of
the correlation lengths associated with the critical phenomena and
the systems ``slows out of equilibrium'', which affects 
the experimental signatures
related to transverse momenta and temperatures.

First order phase transitions are contrarily expected to lead to
large fluctuations due to droplet formation \cite{droplets} or more generally
density or temperature fluctuations. These hot droplets will expand and
hadronize in contrast to cold static quark matter droplets 
that may exist in cores
of neutron stars \cite{HPS}.
In case of a first order phase transition relativistic heavy ion
collisions lead to interesting scenarios in which matter is
compressed, heated and undergoes chiral restoration.  If the
subsequent expansion is sufficiently rapid, matter will pass the phase
coexistence curve with little effect and supercool \cite{Jackson,HJ}.  This
suggests the possible formation of ``droplets'' of supercooled chiral
symmetric matter with relatively high baryon and energy densities in a
background of low density broken symmetry matter. These droplets can
persist until the system reachs the spinodal line and then return
rapidly to the now-unique broken symmetry minimum. A large mismatch
in density and energy density seems to be a robust prediction for a
first order transition at large baryon densities.
At high temperatures, which is more relevant for relativistic heavy-ion
collisions (see Fig. (\ref{Eskola})), 
the transition is probably at most weakly first order as discussed above.

Density fluctuations may appear both for a first order phase
transition and for a smooth cross-over.  If the transitions is first
order, matter may supercool and subsequently create fluctuations in a
number of quantities. Density fluctuations in the form of hot spots or
droplets of dense matter with hadronic gas in between is a likely
outcome (see Fig. (\ref{drop})).  Even if the transition is a smooth
cross-over, the resulting soft equation of state has a small sound
speed, $c_s^2=\partial P/\partial\epsilon$. The equation of state
$P(\epsilon)$ has in both cases a flat region that may be hard to
distinguish in a finite systems existing for a short time only.  We do
not know the early non-equilibrium stages of relativistic nuclear
collisions and the resulting initial density fluctuations, hot spots,
etc. If the system becomes thermalized at some stage, then a smaller
$c_s^2$ is likely to allow for larger density fluctuations since the
pressure difference is smaller. 
Furthermore, in the subsequent
expansion the density fluctuations are not equilibrated as fast when
$c_s^2$ is small because the pressure differences, that drive the
differential expansion, are small. 
The dissipation of an initial density fluctuations can be estimated by a
stability analysis \cite{BFBSZ}. Linearizing the hydrodynamic equations
in small fluctuations around the Bjorken scaling solution,
an entropy fluctuation is typically damped by a factor 
(see Appendix A for details)
\bea
  \frac{\delta S_{final}}{\delta S_{initial}} \simeq 
  \left(\frac{\tau_0}{\tau_f} \right)^{|Re[\lambda_\pm]|} \,. \label{diss}
\eea
Here, a typical formation time is $\tau_0\simeq 1$~fm/c and freezeout
time $\tau_f\simeq 8$~fm/c as extracted from HBT studies \cite{HBT}.
The eigenvalues $\lambda_\pm$ depend on the sound speed and the wave length
of the rapidity fluctuations. As described in more detail in 
Appendix A, one of the eigenvalues are small and vanish for $c_s=0$.
The resulting suppression of an initial density fluctuations during
expansion is typically smaller than a factor 0.5.  If density
fluctuations are enhanced initially due to a softening of the equation
of state due to smooth cross over, then they will largely be preserved
later on. Yet, such fluctuations will be smaller than for a true first
order transition forming supercooled droplets.

Let us assume that hadrons emerge from a collection of
density fluctuations or droplets with 
a Boltzmann distribution with temperature $T$ and from a more or less
continous background obeying approximate Bjorken scaling.
The resulting particle distribution is
\be
 \frac{dN}{dyd^2p_t} \propto  \sum_i f_i\, e^{-m_t\cosh(y-\eta_i)/T} 
     \, +\, {\rm background}\,. \label{dNdy}
\ee
Here, $y$ is the particle rapidity and $p_t$ its transverse momentum,  
$f_i$ is the number of particles
hadronizing from each droplet {\it i}, and  
\be
\eta_i=\frac{1}{2}\log\frac{t_i+z_i}{t_i-z_i}=\frac{1}{2}\log
 \frac{1+v_i}{1-v_i}
  \, 
\ee
is the rapidity of droplet {\it i}. The size, number and
separation between droplets or density fluctuations 
will depend on the violent initial conditions.
Between droplets a relatively continuous background of hadrons
is expected in coexistence.
In (\ref{dNdy}) the droplet is assumed not to expand internally
neither longitudinally nor transversely. If it does expand, the
emerging hadrons will have a wider distribution of rapidities
which will be harder to distinguish from the background.

When $m_t/T\gg 1$, we can approximate
$\cosh(y-\eta_i)\simeq 1+\frac{1}{2}(y-\eta_i)^2$ in Eq.(\ref{dNdy}). 
The Boltzmann factor determines the
width of the droplet rapidity distribution as $\sim \sqrt{T/m_t}$. 
The rapidity distribution will display fluctuations
in rapidity event by event
when the droplets are separated by rapidities larger than
$|\eta_i-\eta_j|\ga \sqrt{T/m_t}$. If they are evenly distributed
by smaller rapidity differences, the resulting rapidity distribution
(\ref{dNdy}) will appear flat.

The droplets are separated in rapidity by $|\eta_i-\eta_j|\sim \Delta
z/\tau_0$, where $\Delta z$ is the correlation length in the dense and
hot mixed phase and $\tau_0$ is the invariant time after collision at
which the droplets form. Assuming that $\Delta z\sim 1$fm --- a
typical hadronic scale --- and that the droplets form very early
$\tau_0\la 1$fm/c, we find that indeed $|\eta_i-\eta_j|\ga
\sqrt{T/m_t}$ even for the light pions.  If strong transverse flow is
present in the source, the droplets may also move in a transverse
direction. In that case the distribution in $p_t$ may be non-thermal
and azimuthally asymmetric.

Even if the transition is not first order, fluctuations may still
occur in the matter that undergoes a transition.  The fluctuations may
be in density, chiral symmetry \cite{DCC}, strangeness, or other
quantities and show up in the associated particle multiplicities.  The
``anomalous'' fluctuations depend not only on the type and order of
the transition, but also on the speed by which the collision zone goes
through the transition, the degree of equilibrium, the subsequent
hadronization process, the amount of rescatterings between
hadronization and freezeout, etc. It may be that any sign of the
transition is smeared out and erased before freezeout. 
That no anomalous event-by-event fluctuations have been
found at CERN \cite{BH} within experimental accuracy indicate that
no transition took place or that the signals were erased before freezeout.
Whether they remain at RHIC is yet to be
discovered and we shall provide some tools for the analysis in the
following sections.

\section{Multiplicity Fluctuations in Relativistic Heavy-Ion Collisions}

\begin{figure}
\centerline{\psfig{figure=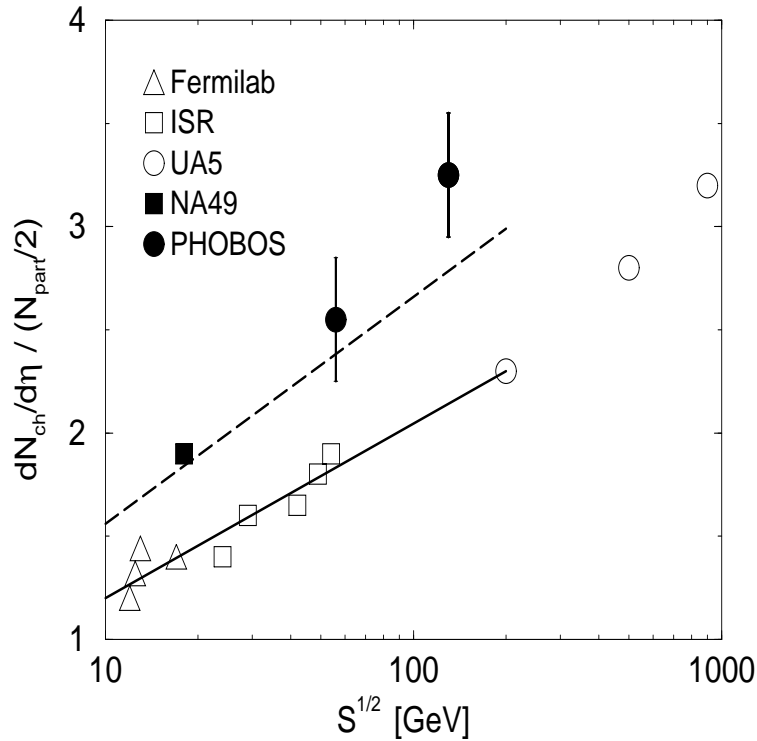,width=11cm,height=11cm,angle=0}}
\caption{Charged particle density per participant 
at midrapidity are shown vs. c.m. energy for
$pp$ and $\bar{p}p$ data \protect\cite{Fermilab,Whitmore,UA5} 
with open symbols. Full curve presents a 
linear fit to $pp$ and $\bar{p}p$ data up
to RHIC energies.
Pb+Pb data from NA49 \protect\cite{NA49} and Au+Au data from PHOBOS 
\protect\cite{PHOBOS} per participant
are shown with filled symbols, and exceed the $pp$ and ${\bar p}p$ by
ca. 30\% (dashed curve).}
\label{s} 
\end{figure}

    In order to be able to extract new physics associated with
fluctuations, it is necessary to understand the role of expected
statistical fluctuations.  Our aim here is to study the sources of
these fluctuations in collisions.  As we shall see, the current NA49
data (see Fig. (\ref{oNA49})) can be essentially understood on the
basis of straightforward statistical arguments.  Expected sources of
fluctuations include impact parameter fluctuations, fluctuations in
the number of primary collisions and in the results of such
collisions, fluctuations in the relative orientation during the
collision of deformed nuclei~\cite{Aa}, effects of rescattering of
secondaries, and QCD color fluctuations.  Since fluctuations in
collisions are sensitive to the amount of rescattering of secondaries
taking place, we discuss in detail two limiting cases, the participant
or ``wounded nucleon model'' (WNM) \cite{WNM}, in which one assumes
that particle production occurs in the individual participant nucleons
and rescattering of secondaries is ignored, and the thermal limit in
which scatterings bring the system into local thermal equilibrium.

Data at AGS, SPS and RHIC energies show that multiplicities are
enhanced by $\sim$30\% in central collisions between heavy
($A\simeq 200$) nuclei as compared to
the WNM prediction (see Fig. (\ref{s})). Whether rescatterings
increase relative fluctuations through greater production of
multiplicity, transverse momenta, etc., or decrease fluctuations by
involving a greater number of degrees of freedom, is not immediately
obvious \cite{BFS,PRL91}. Rescatterings probably increase both the
average multiplicity and its variance but whether the relative amount
of fluctuations are increased is model dependent.  It has even been
found in relativistic heavy ion collisions that the multiplicity
fluctuations increase in the first few rescattering but then decrease
again as the thermal limit is approached. VENUS
simulations~\cite{werner} showed that rescattering had negligible
effects on transverse energy fluctuations.

We first review known multiplicities and fluctuations in the 
basic $pp$ collisions, go on to study nucleus-nucleus
collisions, and finally show in a simple model how 
phase transitions are capable of producing very significant
fluctuations in particle multiplicities.

\subsection{Charged particle production in $pp$ and $p\bar{p}$ reactions}

\begin{figure}
\centerline{\psfig{figure=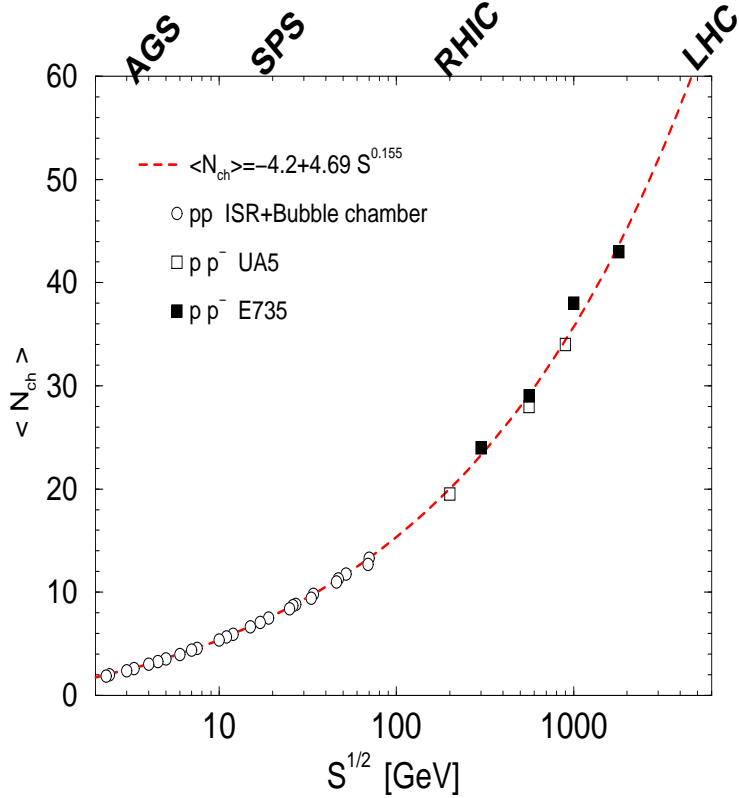,width=11cm,height=11cm,angle=0}}
\caption{The total number of charged particles produced in $pp$
and $p\bar{p}$ collisions vs. cms energy $s^{1/2}$.
Data from bubble chamber \protect\cite{Whitmore},
ISR \protect\cite{Boggild}, UA5 \protect\cite{UA5} and FNAL E735 
\protect\cite{E735}.}
\label{Nchfig} 
\end{figure}

Participant models or WNMs are basically a superposition of NN
collisions. Such models have been studied extensively at these
energies within the last decades at several particle accelerators and
we here give a brief compilation of relevant results.

The average number of charged particles produced in high energy
$pp$ and ultrarelativistic $p\bar{p}$ collisions can be parametrized by
\bea 
  \av{N_{ch}}\simeq -4.2\,+\, 4.69 \left(\frac{s}{{\rm GeV}^2}\right)^{0.155} 
  \,, \label{Nch}
\eea
for cms energies $\sqrt{s}\ga 2$~GeV.
At ultrarelativistic energies the charged particle
production is very similar in $pp$, $pn$ and $p\bar{p}$ collisions and
the parametrization of Eq. (\ref{Nch}) applies
in a wide range of cms energies 2~GeV$\la s^{1/2}\la$~2~TeV
as shown in Fig. (\ref{Nchfig}). At SPS, RHIC and LHC energies,
$\sqrt{s}\simeq 20$, 200, 5000~GeV,
we find $\av{N_{ch}}\simeq 7.3, 20, 60$, respectively.

\begin{figure}
\centerline{\psfig{figure=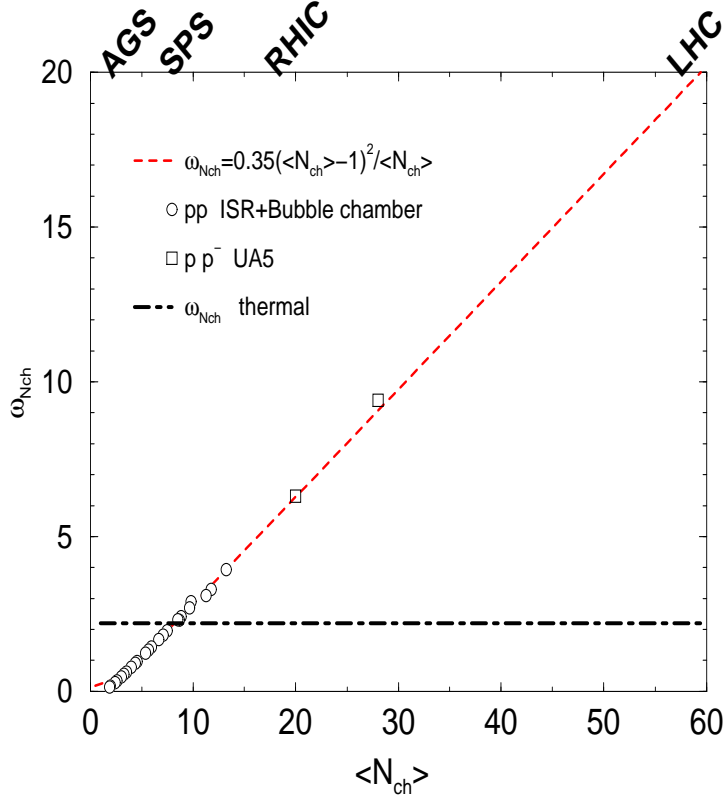,width=11cm,height=11cm,angle=0}}
\caption{The fluctuations in the 
total number of charged particles produced in $pp$
and $p\bar{p}$ collisions from bubble chamber experiments
\protect\cite{Whitmore},
ISR \protect\cite{Boggild}, and UA5 \protect\cite{UA5}.
Note the large difference between pp and thermal fluctuations 
$\omega^{th}_{N_{ch}}\simeq 2.2$ at very high
energies and the ``accidental'' crossing with $pp$ fluctuations
around SPS energies.}
\label{ochfig} 
\end{figure}

At high energies KNO scaling \cite{KNO} is a good approximation.
KNO scaling implies 
that multiplicity distributions are invariant when scaled
with the average multiplicity. Thus all moments scale like
\bea
  \av{N_{ch}^q} \simeq c_q \av{N_{ch}}^q \,, \label{och}
\eea
at high energies where $c_q$ are constants independent of collision
energy. The fluctuations, 
\bea
   \omega_N\equiv (\av{N^2}-\av{N}^2)/\av{N} \,,
\eea
therefore scale with average multiplicity, $\av{N}$, and therefore 
increase with collision energy as in Eq. (\ref{Nch}).
The fluctuations in the charged particle multiplicity can
be parametrized rather accurately by
\bea
  \omega_{N_{ch}} \simeq 0.35 \frac{(\av{N_{ch}}-1)^2}{\av{N_{ch}}}
\eea
as shown in Fig. (\ref{ochfig}) for $pp$ and $p\bar{p}$ collisions
in the same wide range of energies.
At the very high energies breakdown of KNO scaling has been observed
in the direction that 
the fluctuations are slightly larger.
At SPS, RHIC and LHC energies
we find $\omega_{N_{ch}}\simeq 2.0, 6.2, 20$, respectively
in $pp$ and $p\bar{p}$ collisions.

In nuclear (AA) collisions the number of participating nucleons
$N_p$ grow with centrality and nuclear mass number $A$.
Therefore the average charged particle multiplicity and variance
grows with $N_p$, whereas the ratio and therefore the
fluctuation $\omega_{N_{ch}}$ is independent of $N_p$, and
equal to the fluctuations in pp collisions. (Other fluctuations
such as impact parameter will be included below.) 
Higher moments of the multiplicity distributions are large
in high energy pp and $p\bar{p}$ collisions due to KNO scaling
but in nuclear collisions such higher moments are
suppressed by factors of $1/N_p$ and are
therefore less interesting than the second moment.
This justifies our detailed analyzes of the variance
(or rms width) of the fluctuations.

\subsection{Fluctuations in the participant model}

In the participant or wounded nucleon models nucleus-nucleus
collisions at high energies are just a superposition of
nucleon-nucleon (NN) interactions.  In peripheral collisions there are
only few NN collisions, the collision zone is small, rescatterings few
and the WNM should therefore apply. For central nuclear collisions,
however, multiple NN scatterings, energy degradation, rescatterings
between produced particles and other effects complicate the particle
production and {\it do enhance the multiplicities by ca. 30\% as seen
in experiment} (see Fig. (\ref{s})). Thermal models may better
describe central collisions as will be investigated afterwards.  Yet,
the WNM provides a simple baseline to compare to, when going from
peripheral towards central collisions.

  Let us first calculate fluctuations in
the participant model. Although the multiplicities are somewhat
underestimated, the measured
multiplicity and transverse energy in
nuclear collisions at AGS and SPS energies are known to 
scale approximately linearly with the number of
participants~\cite{AGS,NA49}. 
In this picture
\bea
  N=\sum_i^{N_p} n_i,
\label{partmult}
\eea
where $N_p$ is the number of participants and $n_i$ is the number of
particles produced in the acceptance by participant $i$.  In the absence of
correlations between $N_p$ and $n$, the average multiplicity is
$\av{N}=\av{N_p}\av{n}$.  For example, NA49 measures charged particles in the
rapidity region $4<y<5.5$ and finds $\av{N}\simeq270$ for central Pb+Pb
collisions.  Finite impact parameters $(b\la3.5$~fm) as well as surface
diffuseness reduce the number of participants from the total number of
nucleons $2A$ to $\av{N_p}\simeq 350$ estimated from Glauber theory; thus
$\av{n}\simeq0.77$.  Squaring Eq.~(\ref{partmult}) and again assuming
no correlations between different wounded nucleon emission,
$\av{n_in_j}=\av{n_i}\av{n_j}$ for $i\ne j$, we find the multiplicity
fluctuations (see Appendix B)
\bea
   \omega_N =  \omega_n + \av{n}\omega_{N_p}, \label{oN}
\eea
where $\omega_N$, $\omega_n$ and $\omega_{N_p}$ are the 
multiplicity fluctuations in the
total number of particles (within the acceptance), in 
each source,  and in the number of sources respectively.

    A major source of multiplicity fluctuations per participant, $\omega_n$,
is the limited acceptance.  While each participant produces $\nu$ charged
particles, only a smaller fraction $f=\av{n}/\av{\nu}$ are accepted.  Without
carrying out a detailed analysis of the acceptance, one can make a simple
statistical estimate assuming that the particles are accepted randomly, in
which case $n$ is binomially distributed with $\sigma(n)=\nu f(1-f)$ for fixed
$\nu$.  Including fluctuations in $\nu$ we obtain, similarly to
Eq.~(\ref{oN}),
\bea
   \omega_n = 1-f + f \omega_\nu  \,. \label{on}
\eea
In nucleon-nucleon 
collisions at SPS energies, the charged particle multiplicity is
$\sim7.3$ and $\omega_\nu\simeq 1.9$ \cite{GamleOle}; as the multiplicity
should be divided between the two colliding nucleons, we obtain
$\av{\nu}\simeq 3.7$ and thus $f=\av{n}/\av{\nu}= 0.21$ 
for the NA49 acceptance.
Consequently, we find from Eq.~(\ref{on}) that 
$\omega_n\simeq 1.2$. The random acceptance assumption can be improved
by correcting for known rapidity correlations in charged particle
production in $pp$ collisions \cite{Boggild,Whitmore}.

Multiplicities generally increase with centrality of the collision.
We will use the term {\it centrality} as impact parameter
$b$ in the collision. It is not a directly measurable quantity but is
closely correlated to the transverse energy produced $E_T$, the
measured energy in the zero degree calorimeter and the total particle
multiplicity $N$ measured in some large rapidity interval. 
The latter is within the WNM approximately proportional
to the number of participating nucleons
\be
 N_{p}({\bf b})= \int_{overlap} \left[ \rho({\bf r}+\frac{\bf b}{2})
  + \rho({\bf r}-\frac{\bf b}{2}) \right] d^3r \,. \label{Npart}
\ee
For sharp sphere nuclei the number of participants drops from
$N_p(b=0)=2A$ in central collisions to $N_p(b=2R)=0$ in grazing collisions.
For realistic nuclei with diffuse surface and with collision probabilities
given by Glauber theory, the number of participants are 5-10\% smaller
in central collisions but slightly larger in peripheral collisions.

    As a consequence of nuclear correlations, which strongly reduce density
fluctuations in the colliding nuclei, the fluctuations $\omega_{N_p(b)}$ in
$N_p$ are very small for fixed impact parameter $b$ \cite{sigfluct}.  Almost
all nucleons in the nuclear overlap volume collide and participate.  [By
contrast, the fluctuations in the number of binary collisions are
non-negligible.] Cross section fluctuations play a small role in the WNM
\cite{sigfluct}.  Fluctuations in the number of participants can arise when
the target nucleus is deformed, since the orientations of the deformation axes
vary from event to event \cite{NA34}.  The fluctuations, $\omega_{N_p}$, in
the number of participants are dominated by the varying impact parameters
selected by the experiment.  In the NA49 experiment, for example, the zero
degree calorimeter selects the 5\% most central collisions, corresponding to
impact parameters smaller than a centrality cut on impact parameter,
$b_c\simeq 3.5$ fm.  We have
\bea
  \omega_{N_p}\av{N_p} = \frac{1}{\pi b_c^2}
  \int_0^{b_c} d^2b N_p(b)^2 -\av{N_p}^2 \,,
\eea
where $\av{N_p}=(1/\pi b_c^2)\int_0^{b_c} d^2b N_p(b)$.  The number of
participants for a given centrality, calculated in \cite{v2}, can be
approximated by $N_p(b)\simeq N_p(0)(1-b/2R)$ for $0\le b\la 3.5$~fm; thus
\bea
  \omega_{N_p} = \frac{N_p(0)}{18} \left(\frac{b_c}{2R}\right)^2
  \,.\label{oNp}
\eea
For NA49 Pb+Pb collisions with $N_p(0)\simeq 400$ and $(b_c/2R)^2\simeq
5\%$ we find $\omega_{N_p}\simeq 1.1$.  Impact parameter fluctuations are thus
important even for the centrality trigger of NA49.  Varying the centrality cut
or $b_c$ to control such impact parameter fluctuations (\ref{oNp}) should
enable one to extract better any more interesting intrinsic fluctuations.
Recent WA98 analyzes confirm that fluctuations in photons and pions grow
approximately linearly with the centrality cut $(b_c/2R)^2$ \cite{WA98}
as predicted by Eq. (\ref{oNp}).  
The impact parameter fluctuations associated with the range of the
centrality cut, such at total transverse energy or multiplicity,
can therefore be removed. However,
fluctuations in impact parameter may still remain for a given centrality.
The Gaussian
multiplicity distribution found in central collisions changes for minimum bias
to a plateau-like distribution \cite{Aa}.

\begin{figure}
\centerline{\psfig{figure=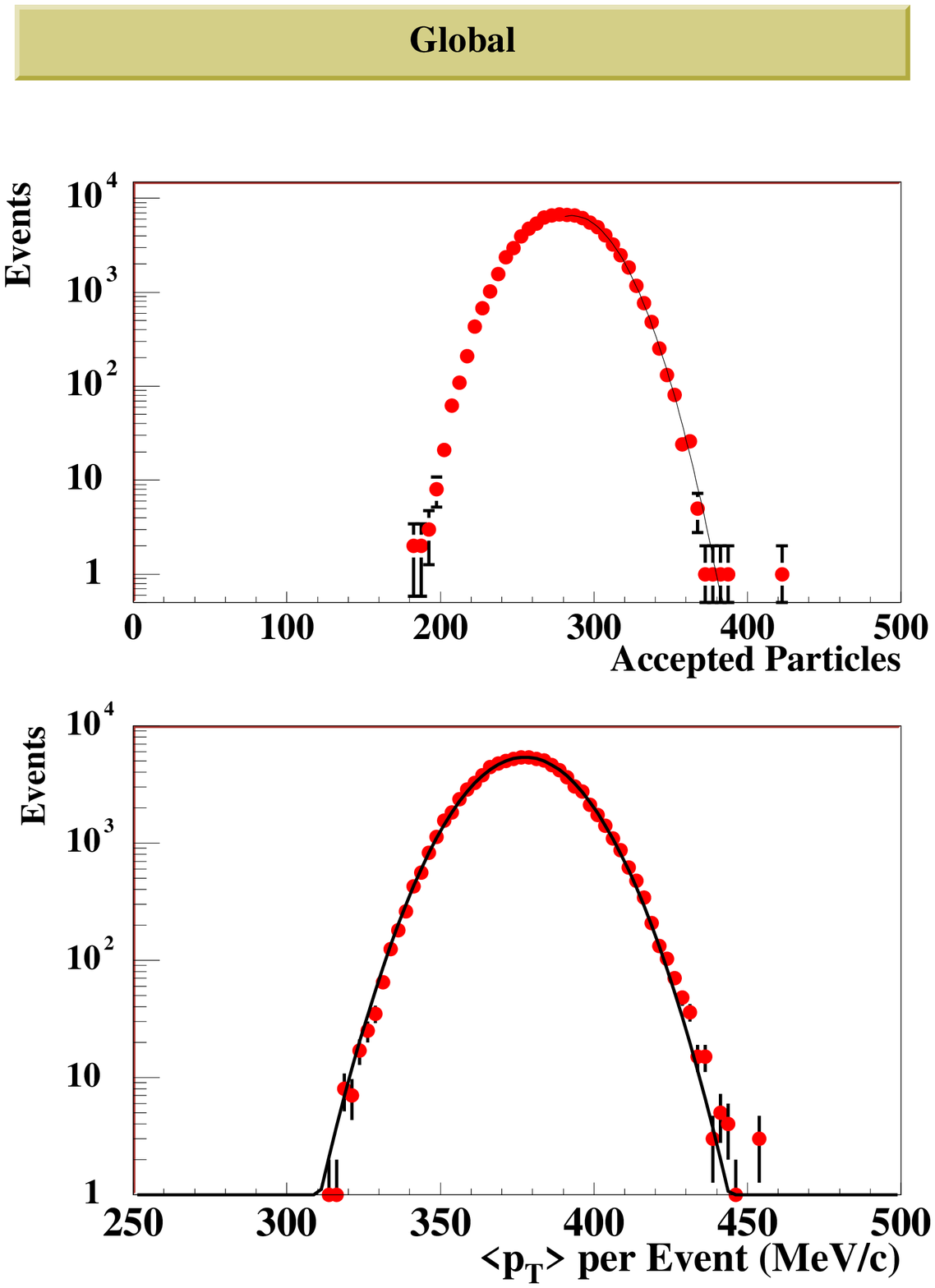,width=13cm,height=21cm,angle=0}}
\caption{Event-by-event fluctuations of multiplicity
(top) and $p_t$ (bottom) measured by NA49 in central Pb+Pb collisions
at the SPS \protect\cite{NA49}.}
\label{oNA49} 
\end{figure}

    Calculating $\omega_N$ for the NA49 parameters, we find from Eq.
(\ref{oN}), $\omega_N\simeq 1.2+(0.77)(1.1)=2.0$, in good agreement with
experiment, which measures a multiplicity distribution $\propto
\exp[-(N-\av{N})^2/2\av{N}\omega_N^{exp}]$, where $\omega_N^{exp}$ is 
$2.01$ \cite{NA49} (see Fig. (\ref{oNA49})).

\subsection{Fluctuations in the thermal model}

    Let us now consider, in the opposite limit of strong rescattering,
fluctuations in thermal models.  In a gas in equilibrium, the mean number of
particles per bosonic mode $n_a$ is given by
\bea
  \langle n_a\rangle = \left(\exp{(E_a/T)}-1\right)^{-1} \,, \label{f}
\eea
with fluctuations
\bea
  \omega_{n_a} = 1+\langle n_a \rangle \,.
\eea
The total fluctuation in the multiplicity, $N=\sum_a n_a$, is
\bea
 \omega^{BE}_N = 1 + \sum_a\langle n_a\rangle^2/\sum_a\langle n_a\rangle .
 \label{oBE}
\eea
If the modes are taken to be momentum states, 
bosons/fermions have thermal fluctuations, $\omega_N=1\pm\av{n_p^2}/\av{n_p}$
where $n_p=(\exp(\epsilon_p/T)\mp1)^{-1}$ is the boson/fermion distribution
function, which are slightly larger/smaller than those
of Poisson statistics for a Boltzmann 
distribution, $\omega_N=1$. 
The resulting fluctuations are 
$\omega^{BE}_N=\zeta(2)/\zeta(3)=1.37$ for massless bosons as, e.g., gluons.
Massive bosons have smaller fluctuations with, for example, $\omega_\pi=1.11$ 
 \cite{Bertsch} and $\omega_\rho=1.01$ when $T=m_\pi$.
Massless fermions, 
e.g.\ quarks, have $\omega_F=2\zeta(2)/3\zeta(3)\simeq0.91$ 
independent of temperature.  

 Resonances are implicitly included in the WNM fluctuations.  In the
thermal limit resonances are found to increase total multiplicity
fluctuations \cite{SRS,BH}
but decrease, e.g., net charged particle fluctuations
\cite{JK1,AHM,JK2,HJ2}.  In high energy nuclear collisions,
resonance decays such as $\rho\to 2\pi$, $\omega\to 3\pi$, etc., lead
to half or more of the pion multiplicity. Only a small fraction
$r\simeq20-30$\% produce two {\it charged} particles in a thermal
hadron gas \cite{Res,JK1} or in RQMD \cite{RQMD} (see also \cite{HH}). 
Not all of the decay
particles from the same resonance always
fall into the NA49 acceptance, $4<y<5.5$,
and the fraction of pairs
will be smaller; we estimate $r\simeq 0.1$. Including such resonance
fluctuations in the BE fluctuations gives, similarly to Eq. (\ref{oN}),
\bea
 \omega^{BE+R}_N = r\frac{1-r}{1+r} + (1+r)\omega_N^{BE} \,. \label{oBER}
\eea
With $r\simeq 0.1$ we obtain $\omega^{BE+R}_N\simeq 1.3$.   
In \cite{SRS} the estimated effect of
resonances is about twice ours:  $\omega_N\simeq 1.5$, not including impact
parameter fluctuations.

    Fluctuations in the effective collision volume add a further term
$\av{N}\sigma(V)/\av{V}^2$ to $\omega^{BE+R}_N$.  Assuming that the
volume scales with the number of participants,
$\omega_V/\av{V}\simeq\omega_{N_p}/\av{N_p}$, we find from
Eq.~(\ref{oN}) that $\omega_N=\omega^{BE+R}_N+\av{n}\omega_{N_p}\simeq
2.1$, again consistent with the NA49 data.  Because of the similarity
between the magnitudes of the thermal and WNM multiplicity
fluctuations, the present measurements cannot distinguish between
these two limiting pictures.

\subsection{Centrality dependence and degree of thermalization}

It is very unfortunate that the WNM and thermal models predict the
same multiplicity fluctuations in the NA49 acceptance - and that they
agree with the experiment. If the numbers from the two models had been
different and the experimental number in between these two, then one
would have had quantified the degree of thermalization in relativistic
heavy ion collisions.

The similarity of the fluctuation in the thermal and WNM is, however,
a coincidence at SPS energies. As seen from Fig. (\ref{ochfig}) the
fluctuations in $pp$ collisions increase with collision energy and
just happen to cross the thermal fluctuations, $\omega_{thermal}\simeq
2.2$, at SPS energies.\footnote{As discussed below the thermal
fluctuations in positive or negative particles are $\omega_\pm\simeq
1.1$ in a thermal hadron gas. The fluctuation in total charge is
twice that due to overall charge neutrality which relates the
number of positive to negative particles.}

\begin{figure}
\centerline{\psfig{figure=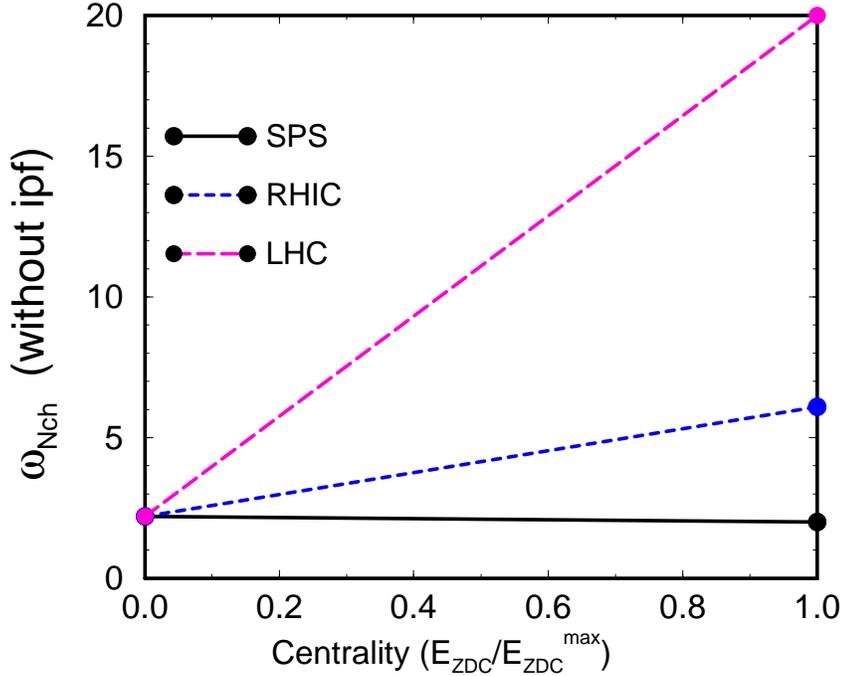,width=12cm,height=10cm,angle=0}}
\caption{The fluctuation in the total number of charged particles
(excluding volume fluctuations) vs. centrality or energy in the
zero degree calorimeter; left are central and right are peripheral
nuclear collisions. 
The curves are linear extrapolations
between the thermal fluctuations, $\omega_{thermal}\simeq 2.2$, in
central collisions to $pp$ fluctuations at SPS, RHIC, and LHC energies, 
$\omega^{pp}$, expected in peripheral collisions within the WNM (see text).
 \label{omegatherm}}
\end{figure}

At RHIC or LHC energies the situation will be much clearer.  Here the
charged particle fluctuations in $pp$ collisions are much larger as
seen in Fig. (\ref{ochfig}), namely $\omega^{pp}_{N_{ch}}=6.5$,~20 at
RHIC and LHC energies respectively. The thermal fluctuations
remain as $\omega_{thermal}\simeq 2.2$. Therefore a dramatic reduction
in event-by-event fluctuations are expected at higher energies at
the nuclear collisions become more central as shown in Fig.
(\ref{omegatherm}).

This can be exploited to define a ``Degree of thermalization''
as the measured fluctuations at a given centrality relative to those
in the thermal and $pp$ limits
\bea
 {\rm Degree\, of\, thermalization} \equiv
\frac{\omega_N^{WNM}-\omega_N^{exp}}{\omega_N^{WNM}-\omega_N^{thermal}} \,,
  \label{therm}
\eea
which ranges from unity in the thermal limit to zero in the WNM.
Whereas both $\omega_N^{WNM}$ and $\omega_N^{exp}$ may depend on
the acceptance the degree of thermalization Eq. (\ref{therm}) 
should not.
Contributions from volume or impact parameter fluctuations may, however,
be centrality
dependent and should therefore be subtracted. Alternatively, the
fluctuations in a ratio, e.g. $N_-/N_+$, should be taken for limited
acceptances.

At RHIC and LHC it should be straight forward to measure the degree
of thermalization as function of centrality. This is interesting on
its own and a necessary requirement for studies of anomalous fluctuations
from a phase transition.

\subsection{Enhanced fluctuations in first order phase transitions} 

First order phase transitions can lead to rather large fluctuations
in physical quantities. Thus, detection of enhanced fluctuations, beyond
the elementary statistical ones considered to this point, could signal the
presence of such a transition.  For example, 
before it became clear that the chiral symmetry restoring phase transition
in hot QCD is not a strong first order phase transition, it was suggested
that matter undergoing a transition
from chirally symmetric to broken chiral symmetry could, when expanding,
supercool and form droplets, resulting in large multiplicity versus rapidity
fluctuations \cite{HJ}.  Let us imagine that $N_D$ droplets fall into the
acceptance, each producing $n$ particles, i.e., $\av{N}=\av{N_D}\av{n}$.  The
corresponding multiplicity fluctuation is (see Appendix B)
\bea
   \omega_N = \omega_n + \av{n} \omega_{N_D} \,. \label{oND}
\eea
    As in Eq.~(\ref{on}), we expect $\omega_n\sim 1$.  However, unlike the
case of participant fluctuations, the second term in (\ref{oND}) can lead
to huge multiplicity fluctuations when only a few droplets fall into the
acceptance; in such a case, $\av{n}$ is large and $\omega_{N_D}$ of order
unity.  The fluctuations from droplets depends on the total number of
droplets, the spread in rapidity of particles from a droplet, $\delta y\sim
\sqrt{T/m_t}$, as well as the experimental acceptance in rapidity, $\Delta y$.
When $\delta y\ll \Delta y$ and the droplets are binomially distributed in
rapidity, $\omega_{N_D}\simeq 1-\Delta y/y_{\rm tot}$, which can be a
significant fraction of unity.

    In the extreme case where none or only one droplet falls into the
acceptance with equal probability, we have $\omega_{N_D}=1/2$ and
$\av{n}=2\av{N}$.  The resulting fluctuation is $\omega_N\simeq\av{N}$, which
is {\it more than two orders of magnitude larger} than the expected value of
order unity as currently measured in NA49. 
It should be said immediately that a much smaller enhancement is realistic
as the transition probably is at most weakly first order and many effects
will smear the signal. Yet, this simple example clearly
demonstrates the importance of event-by-event fluctuations accompanying phase
transitions, and illustrates how monitoring such fluctuations versus
centrality becomes a promising signal, in the upcoming RHIC experiments, for
the onset of a transition.
It is the hope and expectation 
that the higher RHIC energies probe deeper into the QGP phase by
creating higher temperatures and energy densities whereby larger regions
of QGP are produced. The larger event multiplicities should make it possible
to improve on statistics and thereby also the ability to detect
anomalous fluctuations.
The potential for large fluctuations (orders of magnitude) from a
transition makes it worth looking for at RHIC considering the relative
simplicity and accuracy (percents) of multiplicity measurements.

\begin{figure}
\centerline{\psfig{figure=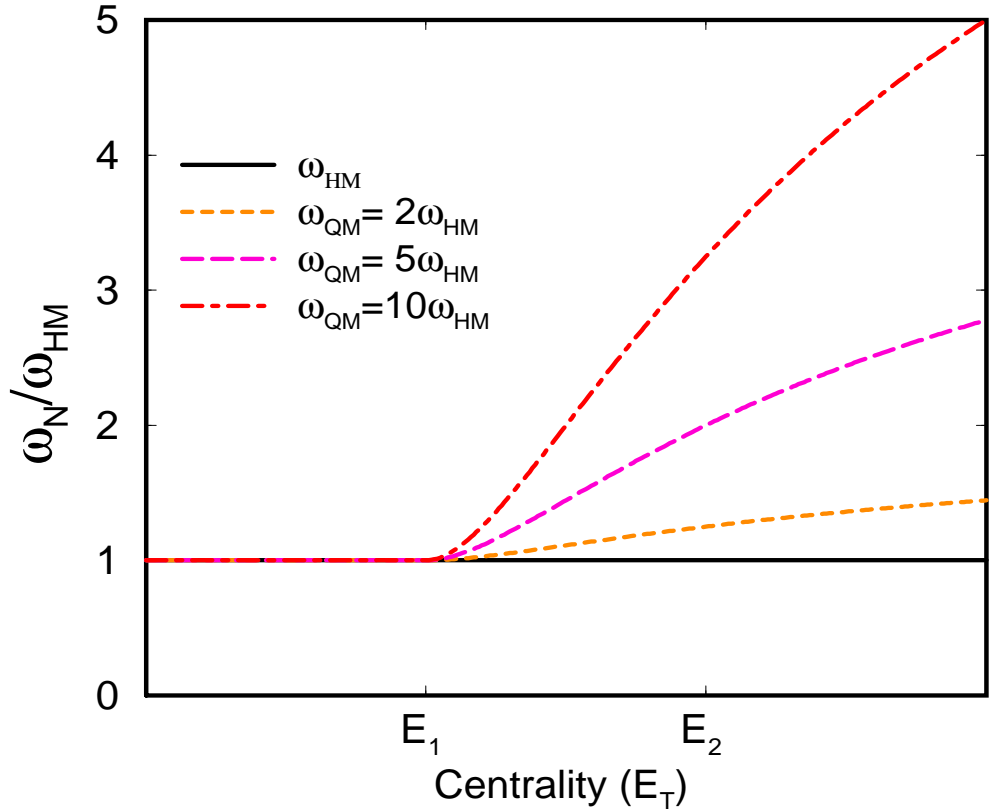,width=15cm,height=12cm,angle=0}}
\vspace{0.2cm}
\caption{Qualitative picture of multiplictity fluctuations 
vs. centrality (total multiplicity or $E_T$).
Anomalous fluctuations appear when a transition to a new state of
matter (QM) starts at centrality $E_1$ (see text).
Curves for different ratios of the fluctuations characterizing the two
states of matter are shown.}
\label{otr} 
\end{figure}

Let us subsequently consider a less extreme model in which a
transition leads to enhanced fluctuations of some kind. 
Assume that the total
multiplicity within the acceptance arises from a normal hadronic
background component ($N_{HG}$) and from a second component
($N_{QGP}$) that has undergone a transition:
\bea
  N = N_{HG}+N_{QGP} \,. \label{NHQ}
\eea
Its average is $\av{N}=\av{N_{HG}}+\av{N_{QGP}}$.  Assuming that
the multiplicity of each of these components is statistically 
independent, the multiplicity fluctuation becomes
\bea
  \omega_N = \omega_{HG} + (\omega_{QGP}-\omega_{HG})
       \frac{\av{N_{QGP}}}{\av{N}} \,. \label{omtr}
\eea
Here, $\omega_{HG}$ is the standard fluctuation in
hadronic matter $\omega_{HG} \simeq 1$. The fluctuations due to the 
component that had experienced a phase transition, $\omega_{QGP}$, depend 
on the type and order of the transition, the speed with which the collision 
zone goes through the transition, the degree of equilibrium, the subsequent 
hadronization process, the number of rescatterings between hadronization and
freezeout, etc.
If  thermal and chemical equilibration eliminate 
all signs of the transition, then $\omega_{QGP} \simeq \omega_{HG}$. 

 The amount of QM and thus $\av{N_{QM}}$
depends on centrality, energy and nuclear masses in
the collision.  For a given centrality the densities vary from zero at
the periphery of the collision zone to a maximum value at the
center. Furthermore, the more central the collision the higher energy
densities are created. The transverse energy, $E_T$, the total multiplicity
and/or the energy in
the zero-degree calorimeter, $E_{ZDC}$, have been found to be
good measures of the centrality of the collision at SPS energies.
Therefore, it would be very interesting to study fluctuations
vs. centrality which are proportional to energy density.  By
varying the binning size for centrality one can also remove
impact parameter fluctuations as discussed above.

If the energy density in the center of the collision zone exceeds the
critical energy density for forming QM at a certain centrality, $E_1$,
then a mixed phase of QM and HM is formed.  At a higher energy
density, where the critical energy density plus the latent heat for
the transition is exceeded, which we shall assume occur at a
centrality $E_2$, then a pure QM phase is produced in the center. 
These quantities will depend on the amount of stopping at a given
centrality, the geometry, $T_c$, etc. 
In the mixed phase $E_1\le E_T<E_2$, the relative amount of QM,
$\av{N_{QM}}/\av{N}$, is proportional to both the volume of the mixed phase.
and the fraction of the volume that is in the QM phase. The latter 
varies in the volume such that it vanishes at HM/QM boundary.

In Fig. (\ref{otr}) a schematic plot of the fluctuations of
Eq. (\ref{omtr}) is shown as function of centrality for various
$\omega_{QM}$. Up to centrality $E_1$ the fluctuations are unchanged.
Above the central overlap zone undergoes the transition to the QM/HM
mixed phase and fluctuations start to grow when
$\omega_{QM}>\omega_{HM}$.  At the higher centrality, $E_2$ the
central overlap zone is in the pure QM phase but the maximum
fluctuations $\omega_{QM}$ are not reached because the surface regions
of the collision zone is still in the HM phase.  On the other hand, if
the hadronization of the QM state is smooth and does not lead to
enhanced fluctuations (i.e. if $\omega_{QGP}=\omega_{HG}$), it cannot
be observed in such a study.

The multiplicity fluctuations can be studied for any kind of
particles, total or ratios. Total multiplicities describe total multiplicities
whereas, e.g. the ratio $\pi^0/(\pi^++\pi^-)$ can reveal fluctuations
in chiral symmetry. The onset and magnitude of such fluctuations
would reveal the symmetry and other properties of the new phase.

\section{Correlations between total and net charge, baryon number
or strangeness}

By a combined analysis of fluctuations in, e.g., positive, negative,
total and net charge as well as ratios, the intrinsic and other 
fluctuations as well as correlations can be extracted and exploited to
reveal interesting physics as will be demonstrated in the following.

\subsection{General analysis of fluctuations and correlations}

Multiplicity fluctuations between various kinds of particles can 
be strongly correlated. As a first example, 
consider the multiplicities of positive and negative pions, $N_+$ and $N_-$, 
in a rapidity interval $\Delta y$ for any relativistic heavy-ion experiment.  
Similar analyzes can be performed for any two kinds of 
distinguishable particles.

The net positive 
charge from the protons in the colliding nuclei is much smaller than the 
total charge produced in an ultrarelativistic heavy-ion
collision.  For example, $\av{N_+}$ exceeds $\av{N_-}$ by 
only $\sim15$\% at in Pb$+$Pb collisions at SPS energies.  The fluctuations 
in the number of positive and negative (or neutral) pions are 
also very similar, $\omega_{N_+}\simeq\omega_{N_-}$.
Charged particle fluctuations have been estimated in thermal as well
as participant nucleon models \cite{BH} including effects of
resonances, acceptance, and impact parameter fluctuations.  By varying
the acceptance and centrality, the degree of thermalization can
actually be determined empirically. Detailed analysis
indicates that the fluctuations in central Pb+Pb 
collisions at the SPS are thermal whereas peripheral collisions are a 
superposition of pp fluctuations \cite{NA49v}.

The fluctuations in the total ($N_{ch}=N_+ + N_-$) and net ($Q=N_+-N_-$) 
charge are defined as \cite{HJ2}
\bea
 && \frac{\av{(N_+\pm N_-)^2}-\av{N_+\pm N_-}^2}{\av{N_++ N_-}}
 = \nonumber\\
 &&\quad \frac{\av{N_+}}{\av{N_{ch}}}\omega_{N_+}
    +\frac{\av{N_-}}{\av{N_{ch}}}\omega_{N_-} \, \pm \, C \,, \label{oNpm}
\eea
where the correlation is given by
\bea
  C = \frac{\av{N_+N_-}-\av{N_+}\av{N_-}}{\av{N_{ch}}/2} \,. \label{C}
\eea
Fluctuations in positive, negative,
total and net charge can be combined to yield 
both the intrinsic fluctuations in the numbers of $N_{\pm}$ and the 
correlations in their production as well as a consistency check. 
These quantities can change as a 
consequence of thermalization and a possible phase transition.

In practice, $\omega_{N_+}\approx \omega_{N_-}$, 
so that the fluctuation in total charge simplifies to
\bea
  \omega_{N_{ch}} &\equiv& \frac{\av{N_{ch}^2}-\av{N_{ch}}^2}{\av{N_{ch}}}
   =\omega_{N_+} + C \,, \label{oNch}
\eea
and that for the net charge becomes 
\bea
  \omega_Q &\equiv& \frac{\av{Q^2}-\av{Q}^2}{\av{N_{ch}}}
   =\omega_{N_+} - C \,. \label{oQ}
\eea

The fluctuation in net charge can 
be related to the fluctuation in the
ratio of positive to negative particles
\bea
 \omega_Q\simeq\av{N_+/N_-}\av{N_{ch}}\omega_{N_-/N_+}/4 \,, \label{oQ+-}
\eea
plus volume
(or impact parameter) fluctuations \cite{JK1,JK2}.  The virtue of this
expression is that volume fluctuations can in principle be extracted
empirically. Alternatively one can vary the centrality bin size or
the acceptance. 
Furthermore, the volume fluctuations for net and total charge are 
proportional to the net ($\av{N_+-N_-}$) and total ($\av{N_++N_-}$) charge
respectively with the same prefactor.
In the following we shall assume that
such ``trivial'' volume fluctuations have been removed.

The analysis has so far been general and Eqs. (\ref{oNpm}-\ref{oQ})
apply to any kind of distinguishable particles, e.g. positive and
negative particles, pions, kaons, baryons, etc. - irrespective of what
phase the system may be in, or whether it is thermal or not.  In the
following, we shall consider thermal equilibrium, which seems to apply
to central collisions between relativistic nuclei, in order to reveal
possible effects on fluctuations of the presence of a quark-gluon
plasma.

\subsection{Charge fluctuations in a thermal hadron gas}

In a thermal hadron gas (HG) as created in relativistic 
in nuclear collisions, pions can be produced either directly or through the 
decay of heavier resonances, $\rho,\ \omega,...$. The resulting
fluctuation in the measured number of pions is
\bea
  \omega_{N_+}=\omega_{N_-} = f_\pi\omega_\pi +f_\rho\omega_\rho+
   f_\omega\omega_\omega + .... \label{of} \, ,
\eea
where $f_r$ is the fraction of measured pions produced 
from the decay of resonance $r$, and $\sum_r f_r=1$. 
These mechanisms are 
assumed to be independent, which is valid in a thermal system. 

The heavier resonances such as $\rho^0, \omega,...$
decay into pairs of $\pi^+\pi^-$ and thus lead to a correlation
\bea
 C^{HG} = \frac{1}{3}f_\rho +f_\omega + .... \,. \label{CHG}
\eea
Resonances reduce the fluctuations in net charge in a
HG in chemical equilibrium at temperature $T=170$~MeV and
baryon chemical potential $\mu_b=270$~MeV and strangeness
chemical potential $\mu_s=74$~MeV to $\omega_Q=0.70$ \cite{JK1,SRS}.
In \cite{AHM} the value $\omega_Q=0.70$ is found.

In addition, overall charge conservation reduces fluctuations in net
charge when the acceptance is large and thus increases correlations as
will be discussed below.

\subsection{Charge fluctuations in a quark-gluon plasma}

A phase transition to the QGP
can alter both fluctuations and correlations in the 
production of charged pions.  To the extent that these effects are not 
eliminated by subsequent thermalization of the HG, they may remain as 
observable remnants of the QGP phase.  
As shown in Refs.\,\cite{AHM,JK2}, net charge fluctuations in a plasma 
of {\it u,\ d} quarks and gluons are reduced partly due to the intrinsically 
smaller quark charge and partly due to correlations from gluons
\bea
 \omega_Q  = \frac{\av{N_q}}{\av{N_{ch}}} \omega_F  
\frac{1}{N_f} \sum_{f=u,d,...}^{N_f} q_f^2 
  \,, \label{oq}
\eea
where $N_f$ is the number of quark flavors, $q_f$ their charges, and
$N_q$ the number of quarks.
The total number of charged particles (but not the net charge) can 
be altered by the ultimate hadronization of the QGP. 
Assuming a pion gas as the final state, this effect can 
be estimated by equating the entropy of all pions to the entropy of the 
quarks and gluons.  
Since 2/3 of all pions are charged and since the entropy 
per fermion is 7/6 times the entropy per boson in a two-flavor QGP
\bea
  \av{N_{ch}} \simeq \frac{2}{3}(\av{N_g} + \frac{7}{6}\av{N_q}) \, ,
 \label{oq2}
\eea
where the number of gluons is $\av{N_g}=(16/9N_f)\av{N_q}$.
Inserting this result in (\ref{oq}), we see that the resulting fluctuations 
are $\omega_Q=0.18$ in a two-flavor QGP 
(and $\omega_Q=0.12$ for three flavors). 
As pointed out in \cite{JK2}, lattice results give 
$\omega_Q \simeq 0.25$. \footnote{It is amusing to note that this number 
gives a very bad (i.e., negative) estimate for $\av{N_g}/\av{N_q}$
in Eq. (\ref{oq2}).}
However, according to \cite{Res} a substantial fraction of the
pions are decay products from the HG, and the
entropy of the HG exceed that of a pion gas by a
factor $1.75-1.8$. As described in \cite{AHM}
the net charge fluctuations
should be increased by this factor in the QGP, i.e.
$\omega_Q\simeq0.33$ in a two-flavor QGP, whereas it remains similar in the
HG, $\omega_Q\simeq0.6$.

The above models are all grand canonical ones, i.e. no net 
charge conservation, as opposed to microcanonical models that now will
be discussed.
If the high density phase is initially dominated by gluons with 
quarks produced only at a later stage of the expansion by gluon fusion, the 
production of positively and negatively charged quarks will be strongly 
correlated on sufficiently small rapidity scales.  
An increased
entropy density in the collisions volume will lead to
enhanced multiplicity as compared to a standard hadronic scenario if
total entropy is conserved.  The associated particle production must
conserve net charge on large rapidity scales ($\Delta y\ga 1$) due to
causality because fields cannot communicate over large distances and
therefore must conserve charge within the ``event horizon''. Therefore
the net charge, $N_{ch}$, is approximately conserved whereas the
total charge, $Q$, increase by an amount proportional to the
additional entropy produced. 
If
the entropy density increases from $s_{HG}$ to $s_{QGP}$ going from a HG 
to QGP without additional net charge production, fluctuations 
in net charge will be reduced correspondingly,
\bea
   \omega_Q^{QGP}\simeq \frac{s_{HG}}{s_{QGP}} \omega_Q^{HG}.
\eea
The resulting 
fluctuation in net charge is necessarily {\it smaller} than that from thermal 
quark production as given by Eq.\,(\ref{oq}).  A similar phenomenon occurs 
in string models where particle 
production by string breaking and $q\bar{q}$ pair production results in 
flavor and charge correlations on a small rapidity scale \cite{Boggild}.

If droplets or density fluctuations appear, they are expected not to
produce additional net charge. Consequently, the net charge
fluctuations should still vanish $\omega_Q\simeq 0$ whereas
$\omega_{ch}\simeq2\omega_{N_+}\sim2\omega_{QGP}$.

The {\it strangeness} fluctuation in kaons $K^\pm$ might seem less
interesting at first sight since strangeness is not suppressed in the
QGP: The strangeness per kaon is unity, and the total number of kaons
is equal to the number of strange quarks.  However, if strange quarks
are produced at a late stage in the expansion of a fluid initially
dominated by gluons, the net strangeness will again be greatly reduced
on sufficiently small rapidity scale.  Consequently, fluctuations in
net/total strangeness would be reduced/enhanced.

The {\it baryon number} fluctuations have been estimated in a thermal
model \cite{AHM} in a grand canonical model. 
It is, however, not known how possible variations in baryon stopping
event-by-event and subsequent diffusion and annihilation of the
baryons and antibaryons in the hadronic phase affect these results.
If only charged particles are detected, but not 
$K^0$, $\bar{K}^0$, neutrons and antineutrons,
the fluctuations have smaller correlations 
as compared to the total and net strangeness or baryon number.

\subsection{Total charge conservation}

Total charge conservation is important when the acceptance
$\Delta y$ is a non-negligible fraction of the total rapidity.  It
reduces the fluctuations in the net charge as calculated within the
canonical ensemble, Eqs.\,(\ref{of}-\ref{oq2}).  If the total
positive charge (which is exactly equal to the total negative charge plus the
incoming nuclear charges) is
randomly distributed, the resulting fluctuations are smaller than the
intrinsic ones by a factor $(1-f_{acc})$, where
\bea
   f_{acc}=(N_{ch}^{tot})^{-1} \int_{\Delta y} \frac{dN_{ch}}{dy}dy
\eea
is the acceptance fraction or the probability
that a charged particle falls into the acceptance $\Delta y$ assuming full
${\bf p}_t$ coverage.  Since charged particle rapidity distributions
are peaked near midrapidity, charge conservation effectively kills
fluctuations in the net charge even when $\Delta y$ is substantially
smaller than the laboratory rapidity, $y_{lab} \simeq 6$ (11) 
at SPS (RHIC) energies.  Total charge conservation also has the effect of 
increasing $\omega_{ch}$ towards $2\omega_{N_+}$ according to 
Eq.\,(\ref{oNch}). Similar effects can be 
seen in photon fluctuations when photons are produced in
pairs through $\rho^0 \to 2\gamma$.  In the WA98 experiment,
$\omega_\gamma \simeq 2$ is found after the elimination of volume
fluctuations \cite{WA98}.

On the other hand, if the acceptance $\Delta y$ is too small, particles can 
diffuse in and out of the acceptance
during hadronization and freezeout \cite{AHM}.  Furthermore, 
correlations due to resonance production will disappear when the average
separation in rapidity between decay products exceeds the acceptance.
Each of these effects tends to 
increase all fluctuations towards Poisson
statistics when $\Delta y\la\delta y$, where $\delta y$ 
denotes the rapidity interval that particles diffuse during hadronization,
freezeout and decay. We find approximately
\bea
   \omega_Q^{exp} \simeq \left(\frac{\Delta y}{\Delta y+2\delta y}\omega_Q
  + \frac{2\delta y}{\Delta y+2\delta y} \right) (1-f_{acc}) 
  \,,  \label{oQexp}
\eea
where $\omega_Q$ is the canonical thermal fluctuation of 
Eqs.\,(\ref{CHG},\ref{oq}) and $\omega_Q^{exp}$ is the 
fluctuation corrected for both $\delta y$ and total charge conservation.

\begin{figure}
\centerline{\psfig{figure=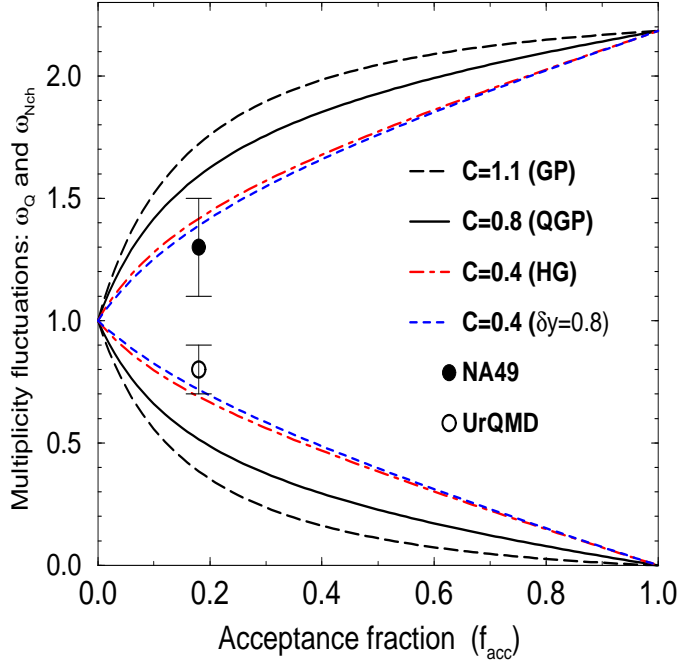,width=10cm,height=10cm,angle=0}}
\vspace{1cm} 
\caption{Acceptance dependence of thermal fluctuations in net charge 
($\omega_Q$ of Eq.\,(\ref{oQexp}), lower curves)
and  total ($\omega_{N_{ch}}$, upper curves).
Correlations increase from a hadron gas ($C\simeq 0.4$), to a QGP
($C\simeq0.8$) and a gluon plasma ($C\simeq1.0$) (see text).
The HG result with a rapidity diffusion of $\delta y=0.8$ is also shown
for comparison to the other curves which use  $\delta y=0.5$.
The large error bar on the NA49 data point is not statistical
but reflects the uncertainty in the subtraction of impact parameter 
fluctuations from fluctuations in charged particles 
\protect\cite{NA49,BH}. The corresponding net charge fluctuation 
predicted by UrQMD \protect\cite{Bleicher} is shown by open circle.
From \protect\cite{HJ2}. } \label{facc} 
\end{figure}

The resulting fluctuations in total and net charge are shown in
Fig.\,(\ref{facc}) assuming $\omega_{N_+}=\omega_\pi\simeq1.1$ and
$\delta y=0.5$. As mentioned above, $f_{acc}$ and $\Delta y$ are
related by the measured charge particle rapidity distributions
\cite{NA49}.  The total charge fluctuations in a HG ($C=0.4$) from
Eq. (\ref{oq2}) agree well with NA49 data \cite{NA49} after
subtraction of residual impact parameter fluctuations. Data on charge
particle ratios, which do not contain impact parameter fluctuations,
will be able to test the net charge fluctuations of Eq.\,(\ref{oQexp})
to higher accuracy.  Predictions from UrQMD are also shown for
comparison \cite{Bleicher}. 
The sensitivity to diffusion is small as seen
in  Fig. (\ref{facc}) where for the fluctuations are
also shown for $\delta y=0.8$ as recently used in \cite{SS}.
The curves in Fig. (\ref{facc}) apply to
RHIC energies as well after scaling $\delta y$ with $\Delta y$.

\section{Fluctuations in particle ratios}

 By taking ratios of particles, e.g. $K/\pi$, $\pi^+/\pi^-$,
$\pi^0/\pi^\pm$, ..., one conveniently removes volume and impact parameter
fluctuations to first approximation.  Simply increasing/decreasing the
volume or centrality, the average number of particles of both species
scales up/down by the same amount and thus cancel in the ratio.

\subsection{$\pi^+/\pi^-$ ratio and entropy production}

 Most particles produced in relativistic nuclear collisions
are pions and they therefore constitute most of the 
number of positive and negatively charged particles.
 The fluctuations in the $\pi^+/\pi^-$ ratio and thus the ratio of
positive and negative particles
are intimately related to the fluctuations in net charge \cite{JK1,JK2}
\bea
 \omega_{N_-/N_+} = \frac{4}{\av{N_{ch}}} \av{N_+/N_-} \, \omega_Q 
   + \omega_{ipf} \,,  \label{o+-}
\eea
where $\omega_{ipf}$ is the impact parameter or volume fluctuations 
and $\omega_Q$ are the net charge
fluctuations as given by Eq. (\ref{oQ+-}).

The $\pi^+/\pi^-$ ratio has been studied in detail in
\cite{JK1}. Resonances such as $\rho,\omega,...$ decaying into two or
three pions correlate the $\pi^+$ and $\pi^-$ production as for
positively and negatively charged particles discussed
above. Consequently, the fluctuation in the $\pi^+/\pi^-$ ratio is
reduced by $\sim30$\% in agreement with NA49 data
\cite{NA49}.

\subsection{$K/\pi$ ratio and strangeness enhancement}

\begin{figure}
\vspace{-2cm}
\centerline{\psfig{figure=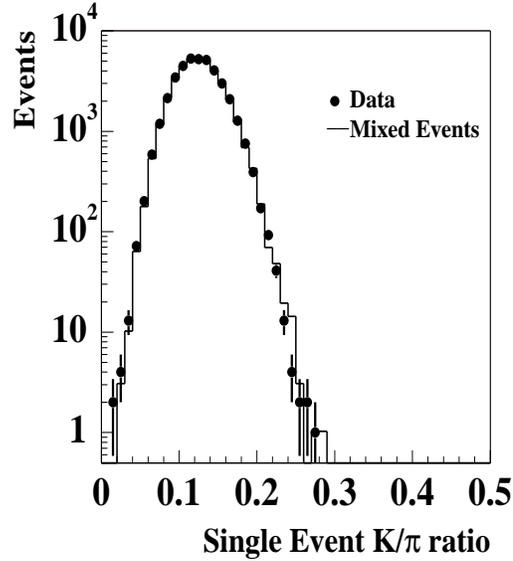,width=15cm,height=25cm,angle=0}}
\vspace{-12cm}
\caption{Event-by-event fluctuations in the $K/\pi$ ratio
measured by NA49 in central Pb+Pb collisions
at the SPS \protect\cite{NA49}.}
\label{oK} 
\end{figure}

  To second order in the fluctuations of the numbers
of K and $\pi$, we have \cite{BH,JK1}
\bea
 \av{K/\pi} =\frac{\av{K}}{\av{\pi}}\left(1+\frac{\omega_\pi}{\av{\pi}} -
               \frac{\av{K\pi}- \av{K}\av{\pi}}{\av{K}\av{\pi}}   \right).
   \label{piK}
\eea
The corresponding fluctuations in $\av{K/\pi}$ are given by
\bea
  D^2\equiv \frac{\omega_{K/\pi}}{\av{K/\pi}} =
  \frac{\omega_K}{\av{K}}+ \frac{\omega_\pi}{\av{\pi}}
     -2\frac{\av{K\pi}- \av{K}\av{\pi}}{\av{K}\av{\pi}} \,.
     \label{D}
\eea
The fluctuations in the kaon to pion ratio is dominated by the
fluctuations in the number of kaons alone.  The third term in Eq.~(\ref{D})
includes correlations between the number of pions and kaons.  It contains a
negative part from volume fluctuations, which removes the
volume fluctuations in $\omega_K$ and $\omega_\pi$ since such fluctuations
cancel in any ratio.  In the NA49 data
\cite{NA49} shown in Fig. (\ref{oK})
the average ratio of charged kaons to charged pions is
$\av{K/\pi}=0.18$ and $\av{\pi}\simeq 200$.  Excluding volume fluctuations, we
take $\omega_K\simeq\omega_\pi\simeq 1.2-1.3$ as discussed above.  The first
two terms in Eq.~(\ref{D}) then yield $D\simeq 0.20-0.21$ 
in good agreement with
preliminary measurements $D=0.23$ \cite{NA49}.  Thus at this stage the data
gives no evidence for correlated production of K and $\pi$, as described by
the final term in Eq.~(\ref{D}), besides volume fluctuations.  The similar
fluctuations in mixed event analyses $D_{mixed}=0.208$ \cite{NA49} confirm
this conclusion.

Strangeness enhancement has been observed in relativistic nuclear
collisions at the SPS.  For example, the number of kaons and therefore
also $\av{K/\pi}$ is increased by a factor of 2-3 in central Pb+Pb
collisions. It would be interesting to study the fluctuations in
strangeness as well. By varying the acceptance one might be able
to gauge the degree of thermalization as discussed above.
The fluctuations in the $K/\pi$ ratio as function of centrality would
in that case reveal whether strangeness enhancement is associated with
thermalization or other mechanisms lie behind. In a plasma of deconfined
quarks strangeness is increased rapidly by $gg\to s\bar{s}$ 
and $q\bar{q}\to s\bar{s}$ 
processes and lead to enhancement of total strangeness $s+\bar{s}$
whereas the net strangeness $s-\bar{s}$ remains zero.
The fluctuations in net and total strangeness will qualitatively behave
like net and total charge, however, with unit strangeness quantum
numbers as compared to the fractional charges.

\subsection{$\pi^0/\pi^\pm$ ratio and chiral symmetry restoration.}

 Fluctuations in neutral relative to charged pions would be a
characteristic signal of chiral symmetry restoration in heavy ion
collisions. If, during expansion and cooling, domains of chiral
condensates gets ``disoriented'' (DCC) \cite{DCC}, anomalous
fluctuations in $\pi^0/\pi^\pm$ ratios could results if DCC domains
are large.  For single DCC domain the probability distribution of
ratios $d=\pi^0/(\pi^0+\pi^++\pi^-)$ is $P(d)=1/\sqrt{2d}$ with mean
$\av{d}=1/3$ and fluctuation $\omega_d=4/15$, i.e. much larger than
ordinary fluctuations in such ratios (see Eq. (\ref{D})) which
decrease inversely with the number of pions.

Neutral pions are much harder to measure than charged pions but with
respect to fluctuations, it suffices to measure the charged pions
only.  The anomalous fluctuations in $\pi^0$ due to a DCC are
anti-correlated to $\pi^\pm$, i.e. they are of same magnitude but
opposite sign. A DCC can equally well be searched for in total charge
fluctuations as in the $\pi^0/\pi^\pm$ ratio, except
for the troublesome impact parameter fluctuations.

\subsection{$J/\Psi$ multiplicity correlations and absorption mechanisms}

$J/\Psi$ suppression has been found in relativistic nuclear collisions
\cite{NA50} and it is yet unclear how much is due to absorption on
participant nucleons and produced particles (comovers).  Whether
``anomalous suppression'' is present in the data is one of the most
discussed signals from a hot and dense phase at early times
\cite{NA50}.  It was originally suggested that the formation of a
quark-gluon plasma would destroy the $c\bar{c}$ bound states
\cite{Matsui}.

In relativistic heavy ion collisions very few $J/\Psi$'s are
produced in each collision. Of these only 6.9\% branch into dimuons
that can be detected and so the chance to detect two dimuon pairs
in the same event is very small. Therefore, it will be correspondingly
difficult to measure fluctuations and other higher moments of the number
of $J/\Psi$. 

Another more promising observable is the correlation between the
multiplicities in, e.g., a rapidity interval $\Delta y$ of a
charmonium state $\psi=J/\Psi,\psi',..$ ($N_\psi$) and all particles
($N$) \cite{HJ2}.  
The correlator $\av{NN_\psi} -\av{N}\av{N_\psi}$ also enters
the in the ratio $\psi/N$ (see Eq. (\ref{piK})).  The correlator has
as good statistics as the total number of $\psi$ and it may contain
some very interesting anti-correlations, namely that $\psi$ absorption
grows with multiplicity $N$. The physics behind can be comover
absorption, which grows with comover density, or formation of
quark-gluon plasma, which may lead to both anomalous suppression of
$\psi$ and large multiplicity in $\Delta y$.  Contrarily, direct
Glauber absorption should not depend on the multiplicity of produced
particles $N$ since it is caused by collisions with participating
nucleons.

To quantify this anti-correlation we model the absorption/destruction 
of $\psi$'s by simple Glauber theory
\bea
  \frac{N_\psi}{N_\psi^0}= e^{-\av{\sigma_{c\psi}\rho_c l}} 
   \equiv  e^{-\gamma N/\av{N}} \,,
\eea
where  $N_\psi^0$ is the number of $J/\Psi$'s before comover or anomalous
absorption sets in but after direct Glauber absorption on participant
nucleons.
In Glauber theory the exponent is the absorption cross section times
the absorber density and path length traversed in matter.
The density and therefore also the
exponent is proportional to the multiplicity $N$ with
coefficient 
\bea
   \gamma=-\frac{d\log N_\psi}{d\log N}_{|N=\av{N}}. 
\eea
In a simple comover absorption model
for a system with longitudinal Bjorken scaling, it can be calculated
approximately \cite{HM}
\bea
  \gamma\simeq \sum_c\frac{dN_c}{dy} 
\frac{\av{v_{c\psi}\sigma_{c\psi}}}{4\pi R^2}  \log R/\tau_0 \,,
\eea
where $dN_c/dy$, $\sigma_{c\psi}$, $v_{c\psi}$ and $\tau_0$ are
the comover rapidity density, absorption cross section, relative velocity
and formation time respectively.

On average comover or anomalous absorption is responsible
for a suppression factor $e^{-\gamma}$. It is difficult to determine
because only the total $\psi$ suppression including direct
Glauber absorption on participants is measured.

The anti-correlation is straight forward to calculate when the fluctuations
in the exponent are small (i.e. $\gamma\sqrt{\omega_N/\av{N}}\ll1$). It is
\bea
  \av{NN_\psi} -\av{N}\av{N_\psi} =
 -\gamma \omega_N \av{N_\psi} \,. \label{NNpsi}
\eea
It  is negative and proportional to the amount of comover
and anomalous absorption and
obviously vanishes when the absorption is independent of multiplicity
($\gamma=0$). The anti-correlation can be accurately determined as
the current accuracy in determining $\av{N_\psi}$ is a few percent
(NA50 minimum bias \cite{NA50}) in each $E_T$ bin.

The anti-correlations in Eq. (\ref{NNpsi}) may seem independent of the
rapidity interval. However, if it is less than the typical
relative rapidities between comovers and the $\psi$, the correlations
disappear. Preferrably, the rapidity interval should be of the order
of the typical rapidity fluctuations due to density fluctuations.

The anticorrelations of Eq. (\ref{NNpsi}) quantify the amount of
comover or anomalous absorption and can therefore be exploited to
distinguish between these and direct Glauber absorption mechanisms.
In that respect it is similar to the elliptic flow parameter for
$\psi$ \cite{HM} for the comover absorption part but differs for the
anomalous absorption.

\subsection{Photon fluctuations: thermal emission vs.
$\pi^0\to2\gamma$}

WA98 have measured photon and charged particle
multiplicities and their fluctuations versus
centrality and $E_T$ binning size. As mentioned above 
impact parameter fluctuations are proportional to the  $E_T$ binning size;
the WA98 analysis nicely confirms this, and can subsequently remove
impact parameter fluctuations. The resulting charged particle
multiplicity fluctuations with impact parameter fluctuations 
subtracted, $\omega_N-\av{n}\omega_{N_p}\simeq 1.1-1.2$
were shown in Fig. (\ref{omegatherm}).

The fluctuations in photon multiplicities were found to be almost
twice as large as for charged particles 
$\omega_\gamma-\av{n}\omega_{N_p}\simeq 2.0$.
This has the simple explanation that photons mainly are produced in
$\pi^0\to2\gamma$ decays. The fluctuations are then the {\it double}
of the fluctuations in $\pi^0$ 
to first approximation as seen from Eq. (\ref{oBER}).

If the photons were directly produced from a ``shining'' thermal
fireball one would expect that they would exhibit Bose-Einstein
fluctuations, $\omega_\gamma=\omega_N^{BE}=1.37$ for massless
particles.  In addition the $\pi^0$'s in the hadronic background will
produce photons with $\omega_\gamma=\omega_N^{BE}=2.0$.  The measured
fluctuation in the number of photons will therefore lie between these
two numbers and can be exploited to quantify the amount of thermal
photon emission vs. $\pi^0\to2\gamma$ decay from a hadronic gas
\bea
  \frac{N_\gamma^{thermal}}{N_\gamma^{thermal}+N_\gamma^{\pi^0}} 
  = \frac{2.0-\omega_\gamma^{exp}}{2.0-1.37} \,.
\eea
The impact parameter fluctuations must be subtracted from the measured
photon fluctuations $\omega_\gamma^{exp}$ by, e.g., 
taking the ratio of photons to some other particle with known behavior.

\section{Transverse momentum fluctuations}

Fluctuations in average transverse momentum
were among the first event-by-event analyses studied.
In a series of papers Mr\'owczy\'nski et al.
have studied transverse momentum fluctuations in heavy-ion collisions
with the purpose of studying thermalization and other effects.
Fluctuations in temperature and thus average transverse momentum
event-by-event were studied by a number of people 
\cite{Stodolsky,Shuryak,SRS,Berdnikov}
in connection with critical phenoma relevant if the transition is close
to a critical point.
Experimental analyses by NA49 \cite{NA49,Trainor} reveal that a
careful evaluation of
systematic effects are required before substantial equilibration can
be claimed in central heavy-ion collisions from transverse momentum
fluctuations. They also have found strong correlations between
multiplicity and transverse momentum.


 The total transverse momentum per event
\bea
P_t=\sum_{i=1}^N p_{t,i} \,,
\eea
is very similar to the transverse energy, for which fluctuations have been
studied extensively~\cite{Aa,PRL91}.  The mean transverse momentum and
inverse slopes of distributions generally increase with centrality or
multiplicity.  Assuming that $\alpha\equiv d\log(\av{p_{t}}_N)/d\log N$ is
small, as is the case for pions \cite{NA44slopes}, the average transverse
momentum per particle for given multiplicity $N$ is to leading order
\bea
   \av{p_{t}}_N = \av{p_t}(1+\alpha (N-\av{N})/\av{N}) \,.
\eea
where $\av{p_t}$ is the average over all events of the single particle
transverse momentum.  With this parametrization, the average total transverse
momentum per particle in an event obeys $\av{P_t/N}=\av{p_t}$.  When the
transverse momentum is approximately exponentially distributed with inverse
slope $T$ in a given event, $\av{p_{t,i}}=2T$, and
$\sigma(p_{t,i})=2T^2=\av{p_t}^2/2$. 

The total transverse momentum and also the transverse energy contains
both fluctuations in multiplicity and fluctuations in the individual
particle transverse momenta and energy (see Appendix C).
An interesting quantity is therefore 
the total transverse momentum per particle, $P_t/N$, where the
multiplicity fluctuations are removed to first order although important
correlations remain. 

    The total transverse momentum per particle in an event has fluctuations
\begin{eqnarray}
 \av{N}\sigma(P_t/N) &=& \sigma(p_{t,i}) + \alpha^2 \av{p_t}^2 \omega_N
  +  \av{\frac{1}{N}  \sum_{i\ne j}(p_{t,i}p_{t,j} -\av{p_{t}}^2)}
  \,.  \label{opt}
\end{eqnarray}
The three terms on the right are respectively:

    {\rm i)} The individual fluctuations $\sigma(p_{t,i})=\av{p_{t,i}^2} -
\av{p_{t}}^2$, the main term.  In the NA49 data, $\av{p_t}=377$~MeV and
$\av{N}=270$.  From Eq.~(\ref{opt}) we thus obtain
$(\sigma(P_t/N)^{1/2}/\av{p_t}\simeq1/\sqrt{2\av{N}}=4.3$\%, which accounts
for most of the experimentally measured fluctuation 4.65\% \cite{NA49}.  The
data contains no indication of intrinsic temperature fluctuations in the
collisions.

    {\rm ii)} Effects of correlations between $p_t$ and $N$, which are
suppressed with respect to the first term by a factor $\sim \alpha^2$.
In NA49 the multiplicity of charged particles is mainly that of pions
for which $T\simeq \av{p_{t}}/2$ increases little compared with pp
collisions, and $\alpha\simeq 0.05-0.1$.  Thus, these correlations are
small for the NA49 data.  However, for kaons and protons, $\alpha$ can
be an order of magnitude larger as their distributions are strongly
affected by the flow observed in central collisions \cite{NA44slopes}.

    {\rm iii)} Correlations between transverse momenta of different
particles in the same event.  In the WNM the momenta of particles
originating from the same participant are correlated.  In Lund string
fragmentation, for example, a quark-antiquark pair is produced with
the same $p_t$ but in opposite direction.  The average number of pairs
of hadrons from the same participant is $\av{n(n-1)}$, where $n$ is
the number of particles emitted from the same participant nucleon, and
therefore the latter term in Eq.~(\ref{opt}) becomes
$(\av{n(n-1)}/\av{n}) (\av{p_{t,i}p_{t,j\ne i}}-\av{p_{t}}^2)$. To a
good approximation, $n$ is Poisson distributed, i.e.,
$\av{n(n-1)}/\av{n}=\av{n}$, equal to 0.77 for the NA49 acceptance, so
that this latter term becomes $\simeq (\av{p_{t,i}p_{t,j\ne
i}}-\av{p_{t}}^2)$.  The momentum correlation between two particles
from the same participant is expected to be a small fraction of
$\sigma(p_{t,i})$.


    To quantify the effect of rescatterings, 
the difference between $\av{N}\sigma(P_t/N)$ and $\sigma(p_{t})$ 
has been studied in detail \cite{GazMrow} via the quantity
\bea
 \Phi(p_t) \simeq \sqrt{\av{N}\sigma(P_t/N)} - \sqrt{\sigma(p_{t,i})}
  \label{pij} \,.
\eea
As we see from Eq.~(\ref{opt}), in the applicable limit that the second
and third terms are small,
\bea
    \Phi(p_t) &\simeq& \frac{1}{\sqrt{\sigma(p_{t,i})}}
    \left(\alpha^2 \av{p_t}^2 \omega_N +
             (\av{p_{t,i}p_{t,j\ne i}}-\av{p_{t}}^2)\right)  \,.\nonumber\\
  && \label{phii}
\eea
In the Fritiof model, based on the WNM with no rescatterings between
secondaries, one finds $\Phi(p_t)\simeq 4.5$~MeV.  In the thermal
limit the correlations in Eq.~(\ref{pij}) should vanish for classical
particles but the interference of identical particles (HBT
correlations) contributes to these correlations $\sim 6.5$~MeV
\cite{Mrow}; they are again slightly reduced by resonances. 
The NA49 experimental value, $\Phi(p_t)=5$~MeV
(corrected for two-track resolution) seems to favor the thermal limit
\cite{NA49}.  Note however that with $\alpha\simeq 0.05-0.1$, the
second term on the right side of Eq.~(\ref{phii}) alone leads to $\Phi
\simeq 1-4$~MeV, i.e., the same order of magnitude.  If
$(\av{p_{t,i}p_{t,j\ne i}}-\av{p_{t}}^2)$ is not positive, then one
cannot a priori rule out that the smallness of $\Phi(p_t)$ does not
arise from a cancellation of this term with $\alpha^2 \av{p_t}^2
\omega_N$, rather than from thermalization.

A comparison of the transverse momentum fluctuations of charged
particles to those in mixed events, where correlations thus are removed,
showed a small enhancement of only $0.002\pm0.002$ \cite{NA49}.
It was estimated that Bose effects should enhance this ratio by
1-2\% but that total energy conservation introduces an anticorrelation
that partially cancels the Bose enhancement \cite{SRS,Berdnikov}.
Experimental problems with two-track resolution have also been estimated
to lead to a ratio that is 1-2\% lower.
Consequently, the numbers seem to be compatible.

The covariance matrix between multiplicity and transverse momentum has
been analyzed by NA49 \cite{NA49}. Strong but trivial
correlations is found due to the fact that higher multiplicity gives
larger total transverse momentum event-by-event. This correlation is
removed in the quantity $P_t/N$ and its covariance matrix with 
multiplicity appears diagonal.

\section{Event-by-Event Fluctuations at RHIC}

 The theoretical analysis above leads to a qualitative understanding of
event-by-event fluctuations and speculations on how phase transitions may show
up.  It gives a quantitative description of AGS and SPS data without
the need to invoke new physics.  We shall here look ahead towards RHIC
experiments and attempt to describe how fluctuations may be searched for.

General correlators between all particle species 
should be measured event-by-event, e.g., the ratios \cite{BH}
\bea
  \frac{\av{N_i/N_j}}{\av{N_i}/\av{N_j}} \simeq 
       1+\frac{\omega_{N_j}}{\av{N_j}} 
       - \frac{\av{N_iN_j}-\av{N_i}\av{N_j}}{\av{N_i}\av{N_j}}   \,,
\eea
where $N_{i,j}$ are the multiplities in acceptances $i$ and $j$ of 
any particle. Volume fluctuations are automatically removed in
such ratios, their fluctuations and correlations. If the energy
deposition, transverse energy or momentum are measured, these latter will
have additional fluctuation due to the multiplicity fluctuations as
explained in Appendix C.

More generally we define the multiplicity correlations between
any two bins
\bea 
 \omega_{ij} = \frac{\av{N_iN_j}-\av{N_i}\av{N_j}}{\sqrt{\av{N_i}\av{N_j}}}
 \,, \label{oij}
\eea
also referred to as the covariance.
When $i,j$ refer to two rapidity bins the covariance is also proportional to
the rapidity (auto-)correlation function $C(y_i-y_j)$.

It is instructive to consider first completely random (uncorrelated or 
statistical)
particle emission. For a fixed total multiplicity $N_{Tot}$, the
probability for a particle to end up in bin $i$ is 
$p_i=\av{N_i}/N_{Tot}\simeq \av{E_i}/E_{Tot}$. The distribution is
a simple multinomial distribution for which
\bea
  \omega_{ij} = \left\{ \begin{array}{lrl}
    1-p_i           & , & i=j \\
    -\sqrt{p_ip_j}  & , & i\ne j       \end{array} \right\} \,.  \label{op}
\eea
The $i=j$ result is the well known one for a binomial distribution. The
$i\ne j$ result is negative because particles in different bins are
anti-correlated: more (less) particles in one bin leads to less (more)
in other bins on average due to a fixed total number of particles.

As shown above there are nonstatistical fluctuations due
to various sources: Bose-Einstein fluctuations, resonances, etc., and
--- in particular --- density fluctuations.
As in Eq. (\ref{NHQ}) we assume that the multiplicity consist of
particles from a HM and a QM phase. The covariances in Eq. (\ref{op})
are derived analogously to Eq. (\ref{omtr})
\bea
   \omega_{ij} = \omega_{ij,HM} + (\omega_{ij,QM}-\omega_{ij,HM})
       \frac{\av{N_{i,QM}}}{\av{N_i}} \,, \label{oijQ}
\eea
when $\av{N_i}=\av{N_j}$; when different 
the general formula is a little more complicated.
Now, the hadronic fluctuations $\omega_{ij,HM}$ is of order unity
for $i=j$, smaller for adjacent bins and vanishes or even becomes 
slightly negative according to (\ref{op}) for bins very different
in pseudorapidity or azimuthal angle $\phi$. The QM fluctuations
can be much larger: $\omega_{i,QM}\sim\av{N_{i,QM}}$ (see the discussion after
Eq. (\ref{oND})).
To discriminate the QM fluctuations from the hadronic ones,
Eq. (\ref{oijQ}) requires
\bea
  \av{N_{i,QM}}\ga 
  \sqrt{(\omega_{ij}-\omega_{ij,HM})\av{N_i}} \,. \label{limit}
\eea
The charged particle multiplicity in central $Au+Au$ collisions at RHIC
is $dN_{ch}/d\eta\simeq 500-600$ per unit pseudo-rapidity \cite{PHOBOS}. 
To see a clear increase in fluctuations, say 
$\Delta\omega\equiv \omega_{ij}-\omega_{ij,HM}\sim 1$, a density
fluctuation of only
$\av{N_{i,QM}}\ga\sqrt{N_i}\simeq 25$ particles are required per unit
rapidity corresponding
to a few percent of the average.
By analyzing many events (of the same total multiplicity)
the accuracy by which fluctuations are measured experimentally is 
greatly improved. Generally, $\Delta\omega\sim1/\sqrt{N_{events}}$,
and so fluctuations can in principle be determined with immense
accuracy.

It may be advantageous to correlate bins with the same pseudorapidity
but different azimuthal angles since the hadronic correlations between
these are small whereas QM fluctuations remain.

 No experimental determination of the purely statistical uncertainties
associated with any one-body distribution --- such as multiplicity as
a function of rapidity --- can be performed without {\em measuring and
diagonalizing\/} the correlation matrix $C_{ij} = \langle N_i N_j
\rangle - \langle N_i \rangle \langle N_j \rangle$.  While it is
conventional to assign uncertainties according to the diagonal
elements $M_{ii}$, the correlations in the covariance matrix are
required for a correct error analysis and can also reveal physical
important results.

\section{Summary}

 A phase transitions in high energy nuclear collisions,
whether it is first order or a soft cross-over,
density fluctuations may be expected that show up in rapidity and
multiplicity fluctuations event-by-event. 
 The fluctuations can be
enhanced significantly in case of droplet formation
as compared to that from an ordinary hadronic scenario. 
A combined analyses of, e.g., positive, negative, total and net charge,
allows one to extract the various fluctuations and correlations 
uniquely.
Likewise a
number of other observables as charged and neutral pions, kaons,
photons, $J/\Psi$, etc., and their ratios can show anomalous
correlations and enhancement or suppression of fluctuations.
This clearly demonstrates the importance of
event-by-event fluctuations accompanying phase transitions, and
illustrates how monitoring such fluctuations versus centrality becomes
a promising signal, in the upcoming RHIC experiments, for the onset of
a transition.  The potential for enhanced or suppressed fluctuations
(orders of magnitude) from a transition makes it worth looking for at
RHIC considering the relative simplicity and accuracy of multiplicity
fluctuation measurements.

 An analysis of fluctuations in central Pb+Pb collisions as currently
measured in NA49 does, however, not show any sign of anomalous
fluctuations.  Fluctuations in multiplicity, transverse momentum,
$K/\pi$ and other ratios can be explained by standard statistical
fluctuation and additional impact parameter fluctuations, acceptance
cuts, resonances, thermal fluctuations, etc. This understanding
by ``standard'' physics should be taken as a baseline for
future studies at RHIC and LHC and searches for anomalous fluctuations
and correlations from phase transitions that may show up in
a number of observables.

By varying the centrality one should be able to determine
quantitatively the amount of thermalization in relativistic heavy ion
collisions as defined in Eq. (\ref{therm}) . For peripheral
collisions, where only few rescatterings occur, we expect the
participant model (WNM) to be approximately valid and the degree of
thermalization to be small. For central collisions, where many
rescatterings occur among produced particles, we expect to approach
the thermal limit and the degree of thermalization should be close to
100\%.  At RHIC and LHC energies the fluctuations in the number of
charged particles consequently decrease drastically with centrality
whereas at SPS energies the two limits are accidentally very close.

Event-by-event physics is an important tool to study thermalization
and phase transitions through
anomalous fluctuations and correlations ---
as in rain.

\section*{Acknowledgements}
Thanks to G. Baym and A.D. Jackson for inspiration and collaboration on
some of the work described in this report.
Discussion with S. Voloshin and G. Roland (NA49),
J.J. G\aa rdh\o je and collaborators in NA44 and BRAHMS, T. Nayak(WA98),
J. Bondorf, S. Jeon, V. Koch, and many suggestions for improvement
from an anonymous referee are gratefully acknowledged.

\newpage

\section*{Appendix A: Damping of initial density fluctuations}

Hydrodynamic flow with
Bjorken scaling is stable according to a stability analysis
carried out in \cite{BFBSZ}.
Bt linearing the hydrodynamic equations in small perturbations 
in entropy density $\delta s$ and rapidity
$\delta y$ around the Bjorken scaling solution and
looking for solutions in the form of harmonic perturbations,
$e^{ik\eta}$, the hydrodynamic
equations could be written in matrix form (Eq. A.13 in  \cite{BFBSZ})
\bea
  \tau\frac{\delta}{\delta\tau} \left( \begin{array}{l}
   \delta s/s \\ \delta y \end{array} \right)
   = \left( \begin{array}{lll}
   0         & & -ik \\
   -ikc_s^2  & & -(1-c_s^2)   \end{array} \right) 
   \left( \begin{array}{l}
   \delta s/s \\ \delta y \end{array} \right)
  \,.
\eea
The eigenvalues of the above matrix
\bea
   \lambda_\pm = -\frac{1}{2}(1-c_s^2)\pm 
   \sqrt{\frac{1}{4}(1-c_s^2)^2-c_s^2k^2} \,, \label{lambda}
\eea
always have real negative part for $c_sk\ne 0$ and fluctuations are
therefore damped. For long wave length fluctuations in rapidity and
not too soft equations of state, $c_sk> 1-c_s^2$, the solution is
a damped oscillator. 
Note that the long wave length solution $k=0$ reproduces the Bjorken scaling.

The exact solution for the entropy density fluctuation
\bea
   \frac{\delta s}{s} = c_+e^{\lambda_+\ln(\tau/\tau_0)}
                      + c_-e^{\lambda_-\ln(\tau/\tau_0)} \,,
\eea
is sensitive to the equation of state through $c_s$, 
the initial conditions for the rapidity density
fluctuations (the constants $c_\pm$), and their wave length $k^{-1}$.

At large times the eigenvalue with the largest real part dominates and
\bea
   \frac{\delta s}{s} \propto 
  \left(\frac{\tau_0}{\tau_f} \right)^{|Re[\lambda_\pm]|} \,. \label{damp}
\eea
Here the oscillating factor has been ignored, leaving the power law
fall-off of fluctuations with exponent
\bea
  Min|Re[\lambda_\pm]| =  \frac{1}{2}(1-c_s^2)-
   Re[\sqrt{\frac{1}{4}(1-c_s^2)^2-c_s^2k^2}]
\eea
One notes that {\it density fluctuations are undamped for soft equation of
states ($c_s=0$)}. They are also undamped if their 
wave length is long ($k\simeq0$).

To estimate the resulting damping we take a typical
rapidity fluctuation for a droplet $\delta y\sim\sqrt{T/m_t}\sim1$
discussed above, which corresponds to a wave-number $k\simeq1$.
For an ideal equation of state with sound speed $c_s=1/\sqrt{3}$ the
last term in Eq. (\ref{lambda}) is then either imaginary or small and real,
and the real part of the eigenvalue is dominated by the first term of
Eq. (\ref{lambda}), $Re[\lambda_\pm]\simeq-1/3$.  If we take a typical
formation time $\tau_0\simeq 1$~fm/c and a freezeout time
$\tau_f\simeq 8$~fm/c as extracted from HBT studies \cite{HBT}, the
resulting  suppression of a density fluctuation during expansion is a factor
$\sim8^{-1/3}=0.5$ according to Eq. (\ref{damp}).

\newpage

\section*{Appendix B: Fluctuations in source models}

 As fluctuations for a source model appears again and again (see
Eqs. \ref{oN},\ref{on},\ref{oBER},\ref{omtr}) we shall derive this
simple equation in detail.

We define the fluctuations for any stochastic variable $x$ as
\bea
  \omega_x  =   \frac{\av{x^2}-\av{x}^2}{\av{x}} \,. \label{omegax}
\eea
It is usually of order unity and therefore more convenient than
variances. For a Poisson distribution, $P_N=e^{-\alpha}\alpha^N/N!$,
the fluctuation is $\omega_N=1$. 
For a binomial distribution with tossing probability $p$ the
fluctuation is $\omega_N=1-p$, independent of the number of tosses.
In heavy ion collisions several processes add to fluctuations so
that typically $\omega_N^{exp}\sim 1-2$.
Correlations can in some cases double the fluctuations as, for example,
$\pi^0\to 2\gamma$ doubles the fluctuations in photon multiplicity
and 
net charge conservation doubles the fluctuation in total charge.
Impact parameter fluctuations further increases the total charge
fluctuations to $\omega_{N_{ch}}=3-5$ in peripheral nuclear collisions
\cite{NA49v}.

 Generally, when the multiplicity ($N$) arise from independent sources
$(N_p)$ such as participants, resonances, droplets or whatever,
\bea
  N=\sum_{i=1}^{i=N_p} n_i,\label{partmult2}
\eea
where $n_i$ is the number of
particles produced in source $i$.  In the absence of
correlations between $N_p$ and $n$, the average multiplicity is
$\av{N}=\av{N_p}\av{n}$. Here, $\av{..}$ refer to
averaging over each individual (independent) source as well as the
number of sources. The number of sources vary from event to event
and average is performed over typically $N_{events}\sim100.000$ events
as in NA49 or $N_{events}\sim10^6$ in WA98.

 Squaring Eq.~(\ref{partmult2}) assuming that the source emit particles
independently, i.e.
$\av{n_in_j}=\av{n_i}\av{n_j}$ for $i\ne j$, the square consists
of the diagonal and off-diagonal elements:
\bea
 \av{N^2} = \av{N_p}\av{n_i^2} + \av{N_p(N_p-1)}\av{n_i}^2 \,. \label{N2}
\eea
With (\ref{omegax}) we obtain the multiplicity fluctuations
\bea
   \omega_N = \frac{\av{N^2}-\av{N}^2}{\av{N}}
            = \omega_n + \av{n}\omega_{N_p}, \nonumber
\eea
as in Eq. (\ref{oN}).

\newpage

\section*{Appendix C: Fluctuations in the energy deposited}

Many experiments do not measure individual particle tracks
or multiplicities but instead the energy deposited in 
arrays of detector segments, $E_i$, in a given event.
One could also project the energy transversely
by weighting with the sine of the scattering angle to study fluctuations
in transverse energy \cite{BFS,PRL91,sigfluct,Aa}.
Since particles mostly have relativistic speeds in relativistic heavy-ion
collisions, the transverse energy 
is almost the same as the total transverse momentum in an event. 

The total energy deposited in the event is
\bea
  E_{Tot} = \sum_i^D E_i \,,
\eea
and can be used as a measure of the centrality of the collision.
The energy deposited in each element (or group of elements)
is the sum over the number of particle tracks ($N_i$) hitting detector $i$
of the individual ionization energy of each particle ($\epsilon_i$)
\bea
  E_i = \sum_n^{N_i} \epsilon_n \,.
\eea
The average is: $\av{E_i}=\av{N_i}\av{\epsilon}$.
The energy will approximately be gaussian distributed,
$d\sigma/dE_i \propto \exp(-(E_i-\av{E_i})^2/2\omega_{E_i}\av{E_i})$,
with fluctuations (see Appendix B)
\bea 
 \omega_{E_i} \equiv \frac{{\av{E_i^2}-\av{E_i}^2}}{\av{E_i}} 
              = \omega_\epsilon+\av{\epsilon}\omega_{N_i} \,. \label{oE}
\eea
Here, the fluctuation in ionization energy per particle
\bea
 \frac{\omega_\epsilon}{\av{\epsilon}} = 
 \frac{\av{\epsilon^2}}{\av{\epsilon}^2}-1 \,,
\eea
depends on the typical particle energies in the detector and 
the corresponding ionization energies for the detector type and thickness.
For the BRAHMS detectors we estimate 
$\omega_\epsilon/\av{\epsilon}\simeq 0.3$ \cite{JJG}. This number
will, however, depend on rapidity since the longitudinal velocity
enters the ionization power.
As these are ``trivial'' detector parameter, we shall exclude the
fluctuations $\omega_\epsilon$ in most
analyses and concentrate on the second term in Eq. (\ref{oE})
which is the fluctuations in the number of particles as examined in
detail above.

\end{document}